\documentclass[a4paper,11pt]{article}
\usepackage{jcappub} 

\usepackage{listings}
\usepackage{xcolor} 
\usepackage{float}
\usepackage{tikz}
\usepackage[T1]{fontenc}
\usepackage[utf8]{inputenc} 
\usetikzlibrary{positioning,fit,arrows.meta,matrix,calc,backgrounds}
\usepackage[margin=2.5cm]{geometry}
\usepackage{soul}

\definecolor{codegreen}{rgb}{0,0.6,0}
\definecolor{codegray}{rgb}{0.5,0.5,0.5}
\definecolor{codepurple}{rgb}{0.58,0,0.82}
\definecolor{backcolour}{rgb}{0.95,0.95,0.92}
\lstdefinestyle{mystyle}{
  backgroundcolor=\color{backcolour},   commentstyle=\color{codegreen},
  keywordstyle=\color{magenta},
  numberstyle=\tiny\color{codegray},
  stringstyle=\color{codepurple},
  basicstyle=\ttfamily\footnotesize,
  breakatwhitespace=false,         
  breaklines=true,                 
  captionpos=b,                    
  keepspaces=true,                 
  numbers=left,                    
  numbersep=5pt,                  
  showspaces=false,                
  showstringspaces=false,
  showtabs=false,                  
  tabsize=2
}
\usepackage[capitalise]{cleveref}
\usepackage{xspace}

\pgfdeclarelayer{preview}
\pgfdeclarelayer{frame}
\pgfsetlayers{background,preview,frame,main}

\newcommand{\ECNN}{\texttt{EquiCNN}\xspace}

\title{\boldmath Dark Matter profiles of {\tt in silico} galaxies: deep learning inference} 

\author{Mart\'in de los Rios$^{1,2}$,}
\author{Serafina Di Gioia$^{1,3}$,}
\author{Fabio Iocco$^{4}$ \&}
\author{Roberto Trotta$^{1,5,6}$}

\affiliation{${}^{1}$SISSA, Via Bonomea 265, 34136 Trieste, Italy and INFN Sezione di Trieste}

\affiliation{${}^{2}$Instituto de Astronom\'ia Te\'orica y Experimental, CONICET - UNC, Laprida 854, X5000BGR, C\'ordoba, Argentina}
\affiliation{$^{3}$ICTP, International Centre for Theoretical Physics, Str. Costiera, 11, Trieste, 34151, Italy}
\affiliation{$^{4}$Dipartimento di Fisica ``Ettore Pancini'', Università degli studi di Napoli ``Federico II'' \& INFN Napoli, Complesso Universitario Monte S. Angelo, I-80126 Napoli, Italy}
\affiliation{${}^{5}$Department of Physics, Imperial College London, SW7 2AZ London, UK}
\affiliation{${}^{6}$Italian Research Center on High Performance Computing, Big Data and Quantum Computing}

\emailAdd{mdelosri@sissa.it}
\emailAdd{mdelosrios@unc.edu.ar}

\abstract{
Machine learning has the potential to improve the reconstruction of the dark matter profile of galaxies with respect to traditional methods, like rotation curves. We demonstrate on the simulation suite Illustris--TNG that a steerable equivariant convolutional neural network (CNN) is able to infer the dark matter profiles within and around individual galaxies from photometric and interferometric data, improving on a standard CNN. 
Within the {\tt in silico} environment of the simulations, our architecture is able to capture the dark matter distribution within galaxies without a parametrization of the profile. We perform an interpretability analysis to understand the internal mechanisms of the trained model and the most important data features used to estimate the dark matter profiles. 
The equivariant CNN recovers the dark matter profile of galaxies within the stellar mass range $[10^{10}-10^{12}] M_{\odot}$ with excellent precision and accuracy: the mean squared error is reduced by a factor of $\sim 3$ from its value under the training distribution, demonstrating that the network has learnt from the data features.
While this holds within the controlled --{\tt in silico}-- environment of the simulation, we argue that few additional steps are needed before this method can be reliably applied to galaxies in the real field observations.

}

\begin{document}
\maketitle
\flushbottom

\section{Introduction}

Over the last three decades, the Standard Cosmological Model, known as $\Lambda$CDM, has established itself as the only self-consistent description of the Universe as a whole, built on tested and validated physical theories such as general relativity, quantum field theory, and thermodynamics, able to explain a wide range of detailed astrophysical observations, and predict many others.
While the $\Lambda$CDM model successfully explains the main properties of the observable universe, our understanding of the true nature of two of its main components (dark matter and dark energy) remains incomplete. Moreover, there are apparent tensions between the predictions of the model on some very specific scales and the recent-most observations. 
In particular, there is a persistent discrepancy between the value of the Hubble constant, $H_{0}$, inferred from early Universe probes, like Planck \cite{planck}, and the value obtained from local Universe probes, for example the measurements based on type Ia supernova observations \cite{riess2016,Riess2022}.
Additionally, the analysis of baryon acoustic oscillations (BAO) by the DESI telescope \cite{desi2025} has provided evidence that dark energy might evolve with time, in contrast to the constant cosmological term assumed in the standard $\Lambda$CDM.
Dark energy is primarily required to explain the accelerated expansion of the Universe on cosmological scales.
Dark matter, on the other hand, is supported by evidence across a wide range of scales and observational probes.  
Historically, the first evidence for dark matter came from galaxy clusters, where Zwicky \cite{Zwicky, Sinclair} identified the need for non-luminous matter to explain the observed velocities of galaxies.
Later, a series of seminal papers \cite{Rubin,Rubin2} provided further compelling evidence through the study of the rotation curves of spiral galaxies. 
More recently, analyses of the Milky Way’s rotation curve \cite{Iocco2015,Iocco2017} have also found significant evidence for the 
impossibility to sustain the rotation curve of our own galaxy through the gravity of the visible component alone.
At larger scales, compelling evidence of the existence of dark matter was found by analyzing the power spectra of the CMB anisotropies and the 2-point correlation function of galaxies.
It is also worth mentioning weak lensing analysis which can directly measure galaxy cluster mass through the bending of light coming from background galaxies. These studies consistently show a discrepancy between the total mass compared with the luminous mass, one more time pointing to the existence of a non-luminous matter component. 
A particularly striking example is the Bullet Cluster \cite{Clowe:2006eq} and other merging clusters \cite{Harvey15}. Analyses combining weak and strong lensing, optical photometry, and X-ray emission data reveal a spatial offset between the distribution of hot gas (traced by X-rays) and the mass (traced by lensing), which is best explained by the presence of collisionless dark matter behaving differently than the collisional intracluster gas \cite{Markevitch04,Harvey15}.

Modelling galaxy formation in a cosmological environment is a challenge, owing to its characteristics of a highly non-linear process that involves physics at different scales.
State-of-the-art cosmological simulations solve the equations of gravitational collapse in an expanding universe including gas particles that, through phenomenological recipes, may form stars.
Being able to reproduce large scale structure patterns, as well as galaxies of different morphology, these simulations have become a fundamental tool for modern cosmology and galactic astrophysics.

Historical evidence for Dark Matter emerges in diverse astrophysical systems spanning a large range of scales, and based on different types of observation and systematics, as pointed out in the recent reviews \cite{cirelli24, bertone18}.
Today, the characterization of the dark matter properties and distribution (often intertwined the one with the other) relies on the mining of a huge amount of data, produced on the one hand by real-sky surveys, and on the other by the numerical modeling of the system against which the data are compared  \cite{bigsparc,kurvs, Romeo2020}. Although rotation curves provide one of the most important pieces of evidence for the existence of dark matter, they are biased tracers of the total mass distribution. In \cite{downing}, it is shown that only in a small fraction of galaxies the rotational velocity of gas particles follows the expected circular velocity relation, due to the complexity of galactic dynamics. Moreover, \cite{Delosrios2111} demonstrated that when observational rotation curves (constructed from synthetic data) are analyzed with traditional methods, the inferred dark matter mass does not accurately reproduce the true underlying mass.
As a specific case, \cite{karukes} shows that the rotation curve method fails to recover the inner slope $\gamma$ of the dark matter density profile of the Milky Way.

In this context, machine learning methods represent a useful alternative, as they are sufficiently flexible (given enough data) to overcome potentially fallacious assumptions required by more traditional analysis methods (like spherical symmetry, dynamical equilibrium, lack of inflows and outflows and the lack of recent major and minor mergers \cite{downing}), while being able to analyze all the available data from modern surveys. Although a diverse range of machine learning techniques has already been brought to bear on a variety of problems with significant success \cite{Villanueva21,Delosrios2111,Wu24,fuaeldiego,Arganda24,Cerdeno25,Sarrato25,Krywonos25}, the ensuing results are often difficult to interpret. Given the highly non-linear nature of these models, their interpretability is an intrinsic limitation, and several groups are working on different solutions to overcome this issue and make the methods more trustable and interpretable \cite{lime,delia,carriero}. 
Another important concern is the robustness of the ML model, that is, how sensitive the final results are to the choice of training data and other modelling choices, like hyperparameters and model architecture. Any unphysical features or biases present in the training set can be learned by the model and subsequently propagate into its predictions when applied to real observations. Inadequate hyperparameters tuning can result in poor generalization on real data.
In particular, several machine learning models have been trained on dataset built from cosmological simulations \cite{Delosrios2111,roger,villanueva22}. While these simulations have converged to a point in which they agree in several coarse-grained aspects, there are still significant discrepancies in some of their main predictions, especially in scales where sub-grid physics became important. This behaviour has already been pointed out in several works \cite{Contardo,Jo2025,villaescusa21} where the models trained on one simulation performed poorly when tested on data from others.

The work presented here is built upon the analysis in \cite{Delosrios2111}, which explored the use of convolutional neural networks (CNNs) to estimate dark matter profiles in and around galaxies from photometric and interferometric information, training the CNNs on synthetic galaxies and mock data built within numerical simulations. In this paper, we upgrade and improve the CNN model of \cite{Delosrios2111}, by explicitly incorporating rotational invariance in the plane of the sky into the network architecture and investigate the relative information content among the various data channels we consider. Additionally, we perform a thorough interpretability analysis of the neural network results, in order to understand both strengths and limitations of the approach, in view of the the long term goal of applying this method to actual data.

This paper is organized as follows. In \cref{sec:meth} we present the dataset used for training, validation and testing of our models, as well as introduce the main properties of the steerable CNNs model along with the specific architectures used in the analysis. In \cref{sec:results} we describe the main results of the paper, including the performance of the models when estimating the dark matter profiles and the interpretability of these results. In \cref{sec:discussion} we discuss the main outcome of our work and how it can be used for enhanced understanding of dark matter distribution of real galaxies in the Universe. 
Finally, in \cref{sec:conclusion}, we present our concluding remarks and outline potential directions for future work.

\section{Methodology}
\label{sec:meth}

Building on \cite{Delosrios2111}, we adopt here a model-agnostic, neural network-enabled dark matter profile reconstruction technique, capable of accounting for the complexity of galactic dynamics and recovering a reliable dark matter profile from photometric observations and HI emission data. To this end, we adopt a supervised ML approach, designed to learn a mapping from input features (in our case, the galaxy's simulated photometry and HI emission) to target outputs (in our case, the dark matter profile). These algorithms require a labelled dataset for training, i.e. a dataset that includes both the input data and the corresponding target variables, which is what we describe next. 

\subsection{Simulated Training Data}
\label{sec:dataset}

In our case, the input features consist of photometric images in the $5$ SDSS bands (u,g,r,i \& z)~\footnote{\href{https://www.sdss.org/dr16/imaging/imaging_basics/}{\url{https://www.sdss.org/dr16/imaging/imaging_basics/}}} \cite{Blanton} and the HI emission data cube, while the target output is a vector with the value of the dark matter mass enclosed within $20$ radial bins of increasing distance from the galaxy's centre.
That is to say that the target vector is constructed by associating to each radial bin the spherical DM mass enclosed in a sphere with radius equal to the upper extreme of the corresponding radial bin as shown in \cref{eq:mass}.

\begin{equation} \label{eq:mass}
    M_{i} = M_{DM} \times N(d < r_{i})  \ ; \  i = 1 \dots 20
\end{equation}

where $M_{i}$ is mass enclosed at a radii $r_{i}$, $M_{DM}$ is the dark matter particle mass corresponding to the used simulation and $N$ is the number of dark matter particles at a distance $d$ smaller than $r_{i}$.
It is worth noticing that in this way, our model reconstructs a \textit{spherical} dark matter profile in a \textit{non-parametric} fashion.
Since the dark matter profiles of real galaxies are not directly observable, we rely on simulated galaxies to construct the labelled dataset. We now describe the cosmological hydrodynamical simulation used, the criteria for subhalo selection, and the main tools and procedures involved in building the dataset. 

Cosmological hydrodynamical simulations model the non-linear evolution of dark matter, gas, and stars particles.
Due to their success in reproducing a wide range of observable quantities \cite{Pillepich}, these simulations have become one of the most reliable controlled environments for training and testing machine learning algorithms in astrophysics. 

We use the TNG100-1 hydrodynamical simulation (\cite{Marinacci2018, Naiman2018, Nelson2018, Pillepich2018, Springel2018}) which adopts a $\Lambda$CDM cosmology ($\Omega_{m} = 0.3089$, $\Omega_{b} = 0.0486$, $\Omega_{\Lambda} = 0.6911$, $\sigma_{8} = 0.8159$, $n_{s} = 0.9667$ and $h = 0.6774$ \citep{Planck2016}). The simulation output consists of $100$ snapshots spanning the redshift range $z=(127,0]$, within a periodic box of side length $75 Mpc/h$  $\approx110.7$ Mpc and includes $2 \times 1820^{3}$ resolution elements. This corresponds to a baryonic mass resolution of $1.39 \times 10^{6} M_{\odot}$ and dark matter particle mass of $7.5 \times 10^6 M_\odot$ \citep{RodriguezGomez2019}.

Each snapshot of the simulation is post-processed using the
\texttt{FOF}  (Friends-of-Friends) and \texttt{SUBFIND} algorithms to identify gravitationally bound haloes and subhaloes and to compute their main properties, including the masses of dark matter, gas, and stars, as well as metallicity and star formation rate, among others \footnote{For a complete description of the available data please refer to \href{https://www.tng-project.org/data/docs/specifications}{\url{https://www.tng-project.org/data/docs/specifications}}}.
In addition, for the hydrodynamical runs, the simulation also includes a flag (named \texttt{SubhaloFlag} in the TNG documentation) indicating whether a subhalo can be considered a galaxy.
In what follows, we refer to subhaloes flagged as galaxies by  \texttt{SubhaloFlag} simply as ``galaxies'' for clarity.

\subsection{Subhalo and galaxy selection}
\label{sec:galdataset}
Following \cite{Delosrios2111} we select all the central galaxies of snapshot $99$ (corresponding to $z=0$) with stellar mass between $10^{10}  M_{\odot} \leq M_\star \leq 10^{12}  M_{\odot}$
and star formation rate $\text{SFR} \geq 0.1$ $[M_{\odot} / yr]$ (See \cref{tab:selection_criteria} for a complete description of the selection criteria). We then place each galaxy at a random distance, $D$, uniformly distributed in the range $[10,20]$ Mpc, and with a random inclination angle, $\theta$, with respect to the line of sight (uniformly distributed in the range $[0, \pi/2]$), thus ensuring a sample of galaxies with diverse configurations in their spatial distributions. 

\begin{table}
    \centering
    \begin{tabular}{c|c}
        Property & Criterium for selection \\ \hline
        Simulation snapshot & $99$ $(z = 0)$ \\
        Stellar mass & $10^{10} \; M_\odot \leq M_\star \leq 10^{12} \; M_\odot$ \\
         Star formation rate & $\text{SFR} \geq 0.1$ $M_\odot / \text{yr}$ \\
        Central galaxy & SubhaloID = GroupFirstSub \\
        Cosmological origin & SubhaloFlag = 1
    \end{tabular}
    \caption{Summary of the criteria used in selecting galaxies from the TNG100-1 simulation.}
    \label{tab:selection_criteria}
\end{table}

For each galaxy we compute the fraction of kinetic energy
invested in ordered rotation, $\kappa$, as defined in \cite{Sales2012}:

\begin{equation}\label{eq:kappa}
 \kappa = \frac{K_{rot}}{K} = \frac{1}{K} \sum_{i}^{N_{stars}} \frac{m_{i}}{2} \left( \frac{j_{z,i}}{R_{i}} \right)^{2}
\end{equation}
where $K$ is the total kinetic energy of the stars and $m_{i}$, $j_{z,i}$ and $R_{i}$
represent the mass, the z-component of the angular momentum and the projected radius of each star particle $i$, respectively. 
This ratio measures how rotationally supported is the galaxy, with larger value of $\kappa$ corresponding to more rotationally supported galaxies. As a proxy for their morphological class, galaxies with $\kappa > 0.5$ can be considered spirals while galaxies with $\kappa < 0.5$ can be considered ellipticals.
This procedure yields a dataset consisting of $2972$ independent galaxies, with spiral galaxies ($\kappa > 0.5$) accounting for $\sim 77\%$ of the total dataset, the remaining being elliptical galaxies ($\kappa < 0.5$).

For each galaxy we create $5$ photometric images, and the HI-emission datacube that is summarized to $3$ maps that correspond to the line-of-sight intensity (HI-0), average velocity (HI-1) and velocity dispersion (HI-2)  maps, respectively. Since the simulation provides full information about all dark matter particles gravitationally bound to each galaxy, we can compute the enclosed dark matter mass at 20 radial bins in geometric space, ranging from 1 to 100 kpc\footnote{Owing to the fixed spatial extent of bins, for galaxies whose size is smaller than 100 kpc, the value of bins at radii larger than the galaxy's size is constant.}. These radial mass profiles serve as the target outputs for our machine learning models. 
We have also tested that changing the number of bins, while keeping the maximum radii at $100$ kpc, does not affect qualitatively the results. 

As usual in any machine learning analysis, to avoid overfitting and obtain better generalisation, the data set was randomly split into $70\%$, $25\%$, and $5\%$ to build the training, validation, and testing sets, respectively.
We also normalize both the input (i.e. the photometric and interferometric maps) and the output (i.e. the dark matter mass enclosed at different radius) variables between $0$ and $1$, taking into account the minimum and maximum values of each variable across the training set.

\subsection{Mock data generation}
\label{sec:mockdata}

To generate photometric imaging data and HI emission data cubes that  mimic how these simulated galaxies would appear to real astronomical instruments, we make use of the radiative transfer code \texttt{SKIRT} \footnote{\href{https://github.com/SKIRT/SKIRT9}{https://github.com/SKIRT/SKIRT9}}\cite{skirt0,skirt,skirt9} and the radio simulation package \texttt{MARTINI} \footnote{\href{https://github.com/kyleaoman/martini}{https://github.com/kyleaoman/martini}} \cite{martini0,martini1,martini2} code.
 All the mock observations
are rendered to cover a field of view of $17.2' \times 17.2'$ with a resolution of
$128 \times 128$ pixels.

\texttt{SKIRT} is a 3D Monte Carlo radiative transfer code specifically developed for modeling the stellar emission in astrophysical systems and subsequent light-ray propagation to the observer, including the interaction with interstellar dust, multiple anisotropic scattering, and thermal dust emission. 
This code takes as input a given cut-out of particles from the hydrodynamical simulation and supports different dust models and several spectral energy distributions (SEDs) for the star particles.

\texttt{MARTINI} (Mock Array Radio Telescope Interferometry of the Neutral ISM) is Python package designed to generate synthetic HI (21-cm) emission data cubes from smoothed-particle hydrodynamics (SPH) galaxy simulations. 
It simulates realistic radio observations by incorporating effects such as telescope beam response, spectral resolution, and observational noise. 
We choose the spectral dimension of the data cube (64 channels) and the spectral resolution ($5$ km/s), to match the properties of
the THINGS survey \cite{things} obtained with the VLA radio telescope.

A detailed description of the configuration parameters used for both \texttt{SKIRT} and \texttt{MARTINI} can be found in \cref{sec:skirtmartini}.

\begin{figure}
    \centering
    \includegraphics[width=1\linewidth, trim={0 2.4cm 0 4cm}, clip]{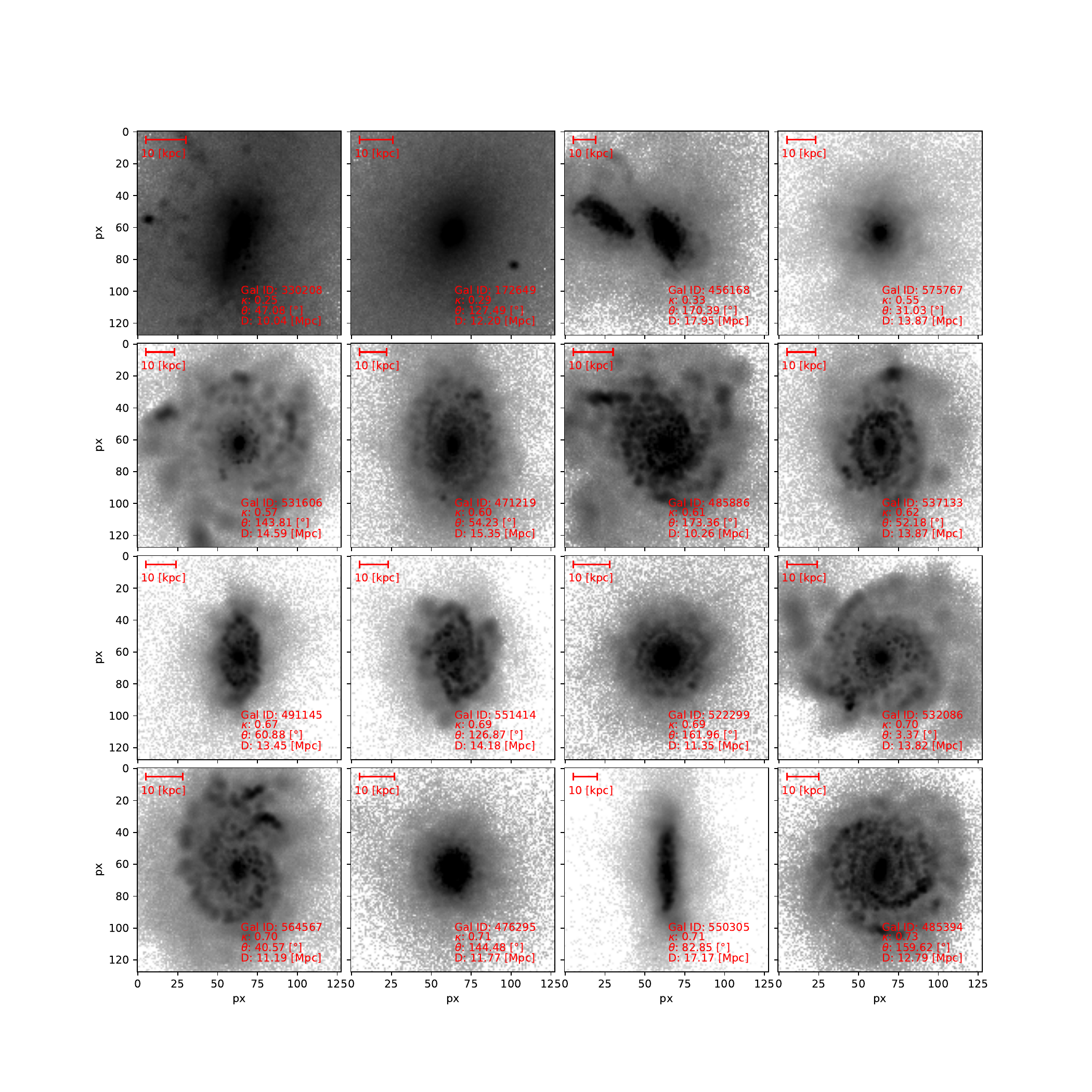}
    \caption{Examples of simulated SDSS images of galaxies in the $g$ band produced with \texttt{SKIRT}. In the bottom right corner we show the corresponding subhalo ID, the fraction of rotational energy $\kappa$ and the distance $D$ and inclination angle $\theta$ of the rendered configuration.}
    \label{fig:sdss_examples}
\end{figure}

\begin{figure}
    \centering
    \includegraphics[width=1\linewidth, trim={0 2.4cm 0 4cm}, clip]
    {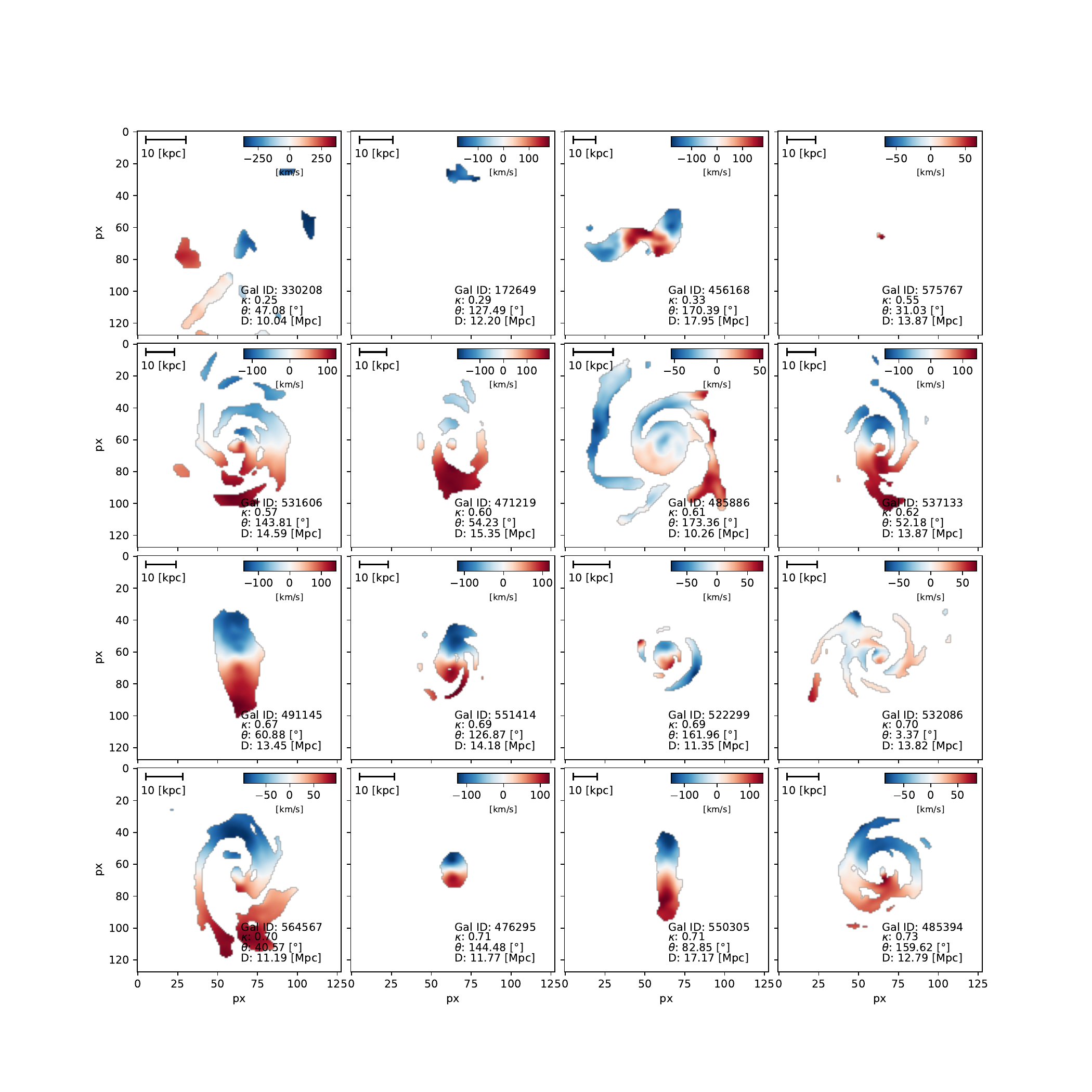}
    \caption{HI first momentum map (HI$-$1) produced with \texttt{MARTINI} for the same simulated galaxies as in Fig.~\ref{fig:sdss_examples}. The colormap represents the relative gas velocity with respect to the mean velocity. We add, in the bottom right corner, the corresponding subhalo ID, the fraction of rotational energy $\kappa$ and the distance $D$ and inclination angle $\theta$ of the rendered configuration.}
    \label{fig:h1_examples}
\end{figure}

In \cref{fig:sdss_examples,fig:h1_examples} we show the synthetic images produced with \texttt{SKIRT} and \texttt{MARTINI} for $16$ random galaxies.

\subsection{Steerable Equivariant Convolutional Neural Networks} 
\label{sec:models}

\begin{figure}[t]
\centering
\begin{tikzpicture}[
  font=\footnotesize,
  node distance=4mm,
  >={Stealth[length=2mm]},
  obox/.style={draw, rounded corners, inner sep=1pt},
  ibox/.style={draw, rounded corners, align=center, minimum height=5.5mm, text width=20mm},
  title/.style={font=\bfseries}
]

\tikzset{ 
   obox/.style={ draw, rounded corners=4pt, very thick, inner sep=2mm, fill=gray!3 
   }, 
   ibox/.style={ draw, rounded corners=2pt, thick, inner xsep=2mm, inner ysep=1.5mm, align=center, font=\footnotesize, fill=white}, 
   title/.style={font=\bfseries}, 
   note/.style={font=\scriptsize, align=left} 
   
} 

\newcommand{\fcPreviewCount}{4} 
\newcommand{\fcPreviewShiftX}{+2mm} 
\newcommand{\fcPreviewShiftY}{+3mm} 
\newcommand{\fcPreviewDraw}{black!35} 
\newcommand{\fcPreviewFill}{gray!3} 

\node[ibox, text width=16mm] (inp) {Input\\$(C,H,W)$\\};

\newcommand{\RtwoBlock}[4]{%
  \begin{pgfonlayer}{main}
    \node[ibox, right=9mm of #1] (relu-#4) {ReLU (e2nn)};
    \node[ibox, above=1mm of relu-#4] (bn-#4) {InnerBN };
    \node[ibox, above=1mm of bn-#4] (conv-#4)  {R2Conv};
    \node[ibox, below=1mm of relu-#4] (drop-#4) {Dropout2d };
    \node[ibox, below=1mm of drop-#4, text width=14mm] (pool-#4) {MaxPool $2{\times}2$\\ };
  \end{pgfonlayer}
  \begin{pgfonlayer}{frame}
    \node[obox, fit=(conv-#4)(pool-#4),text width=22mm, label={[title]above:#2}] (#4) {};
  \end{pgfonlayer}
  
  \begin{pgfonlayer}{background} 
    \foreach \i in {2,1} {%
    \draw[rounded corners=4pt, very thick, draw=\fcPreviewDraw, fill=\fcPreviewFill] 
    ($(#4.north east)-(\i*\fcPreviewShiftX,\i*\fcPreviewShiftY)$) rectangle 
    ($(#4.south west)-(\i*\fcPreviewShiftX,\i*\fcPreviewShiftY)$); }%
  \end{pgfonlayer} 
}

\RtwoBlock{inp}{3 $\times$ Conv. blocks}{Out: $5\times$ regular, $|G|{=}4$}{B1}

\draw[->] (inp.east) -- (B1.west);

\node[ibox, right=10mm of B1, text width=20mm] (gp) {GroupPooling\\};
\draw[->] (B1.east) -- (gp.west);

\node[ibox, below=6mm of gp, text width=15mm] (fl) {Flatten};
\draw[->] (gp.south) -- (fl.north);

\newcommand{\FCblock}[4]{%
  \begin{pgfonlayer}{main}
    \node[ibox, below=12mm of #1]  (fc-#4) {Fully-Connected layer};
    \node[ibox, below=1mm of fc-#4] (bn-#4) {InnerBN (opt)};
    \node[ibox, below=1mm of bn-#4] (relu-#4) {ReLU (e2nn)};
  \end{pgfonlayer}
  \begin{pgfonlayer}{frame}
     \node[obox, fit=(fc-#4)(relu-#4), label={[title]above:#2}] (#4) {};
  \end{pgfonlayer}

  \begin{pgfonlayer}{background} 
    \foreach \i in {4,3,2,1} {%
    \draw[rounded corners=4pt, very thick, draw=\fcPreviewDraw, fill=\fcPreviewFill] 
    ($(#4.north east)-(\i*\fcPreviewShiftX,\i*\fcPreviewShiftY)$) rectangle 
    ($(#4.south west)-(\i*\fcPreviewShiftX,\i*\fcPreviewShiftY)$); }%
  \end{pgfonlayer} 
}

\FCblock{fl}{$5\times$ FC blocks }{Out: $5\times$ regular, $|G|{=}4$}{F1}

\draw[->] (fl.south) -- (F1.north);
\node[ibox, right=4mm of F1, text width=14mm] (output) {output};
\draw[->] (F1) -- (output);

\end{tikzpicture}
\caption{Schematic layout of our equivariant steerable CNN, \ECNN.
C, H and W indicate to the number of channels, and the height and width of the images, in our case $C=8$ and $H=W=128$. Our network has 3 equivariant convolutional blocks, each encompassing one R2Conv layer, as defined in the library \texttt{e2cnn}, one Inner BatchNorm Layer (BN), the ReLU layer, Dropout and MaxPooling layers. The third Equivariant Convolution block is followed by a GroupPooling, a Flatten layer and 5 Fully-Connected (FC) Layers.}

\label{fig:E2CNN}
\end{figure}

Classical convolutional neural networks (CNNs) are by design equivariant to translations of their input, meaning that they guarantee invariance of their feature spaces under a translation of their input. Steerable CNNs represent a generalization of traditional CNNs, with steerable kernels designed to be equivariant under the action of a specific transformation group. In problems with rotational symmetry, a traditional CNN needs to learn rotated versions of the same filter, thus introducing redundant degrees of freedom and reducing generalization to trivial rotations of the same object. By contrast, steerable equivariant CNNs can incorporate in their structure the information about the group transformations acting on the input data and preserve the input data invariance under the action of such groups \cite{cohen2016steerable}.

Among the possible definitions of steerable CNNs, we adopt here the `E2CNN' architecture, designed to be equivariant under all isometries $E(2)$ of the image plane $\mathrm{R}^2$, which include translations, rotations and reflections. 
In particular, our steerable CNN is defined to be equivariant under rotations of the galaxy images and datacubes by $\pi/2$, respecting equivariance under the action of element of the $C_2$ cyclic group.  We chose to consider only equivariance under this symmetry group because rotations of multiples of $\pi/2$ automatically preserve the pixelized structure of a squared image (or datacube), thus simplifying the setup. Furthermore, previous work~\citep{cohen2016group} has shown that equivariant CNNs yield state-of-the-art results on classification tasks even if they enforce equivariance only to
small groups of transformations like rotations by multiples of 90 degrees (for example in the case of digits). Indeed, we demonstrate in Appendix \ref{app:rotation-test} that adding this structural rotational invariance to the network architecture (despite being limited to rotations of multiples of $\pm \pi/2$) improves the generalization capabilities of the model as compared to standard CNNs. Generalization of this setting to groups describing rotation with arbitrary angles will be the subject of future work. 

A schematic representation of the equivariant CNN network, which we call \ECNN, and its components is given in Fig. \ref{fig:E2CNN}. We implement the \ECNN using the standard equivariant convolution layers (R2Conv) defined in the library \texttt{e2cnn} \footnote{\href{https://github.com/QUVA-Lab/e2cnn}{https://github.com/QUVA-Lab/e2cnn}}, following~\citep{weiler2019general}, MR{ and preserving the same number of hidden layers adopted in \cite{Delosrios2111}. The number of irreducible representations to be used in the R2Conv layers and all the hyperparameters (e.g. initial learning rate) were selected via manual tuning based on empirical validation performance.}

\subsection{Training and validation}
\label{sec:trainval}

The \ECNN is trained with an \texttt{MSE} loss function:

\begin{equation} \label{eq:mse}
MSE = \frac{1}{N_{obs}N_{bins}}\sum_{i}^{N_{obs}}\sum_{j}^{N_{bins}} \left[ \log{M_{REAL,j}^{i}} - \log{M_{REC,j}^{i}} \right]^{2}
\end{equation}
where $M_{REAL,j}$ and $M_{REC,j}$ are the real and reconstructed dark matter mass of the $i^{th}$ galaxy inside the $j^{th}$ radial bin. We use the \texttt{ADAM} optimizer with a learning rate of $lr = 0.0006$.
We train for 20 epochs with early stopping patience of $10$, to avoid the model entering the overfitting phase. We adopt the following bootstrap approach for uncertainty quantification: we split the data into training (70\%), validation (20\%) and testing (10\%), with validation and testing sets kept fixed. We then train $20$ \ECNN instances, bootstrapping the training set (i.e., we built $20$ training sets by randomly selecting observations with replacement from the original training set). The number 20 has been chosen to remain conservative: we have observed no variation between 20 and 30 sets in a training test, but did not systematically explore lower repetitions. It is possible that a smaller number would deliver the same results, and make the procedure computationally lighter in the future. At testing time, for each galaxy we have $20$ estimated profiles (one for each boostrapped instance of the model) from which we compute the average reconstruction and its standard deviation, which is a measure of both stochasticity in the training and variability in the fitted network's weights. Our bootstrap uncertainty estimate thus includes both aleatoric and epistemic uncertainties. In order to reduce epistemic uncertainty (i.e., variance in the fitted model parameters) one would need to increase the training size, which is not feasible giving computational constraints at the moment. A cleaner estimate of epistemic uncertainty could be obtained with a Bayesian approach (i.e., Bayesian Neural Networks), which however would require a major change in the current approach. Furthermore, we expect that model misspecification due to out-of-distribution effects when comparing simulations with real data will play an important role should this method be applied to actual observations. This requires dedicated effort to tackle and we therefore leave this estimate to future work. We could have used leave-one-out cross-validation, but this would have been computationally expensive. k-fold cross-validation was not an option due to the small size of the training set.

The \ECNN model also delivers computational advantages: training the 20 instances of the \ECNN model takes less than half an hour on one GPU NVIDIA A100, on Google Colab, for our dataset. Training the 20 instances of the \ECNN model takes only 10\% longer than training of 20 instances of the standard CNN (defined in Appendix~\ref{app:rotation-test}) on the same data. By contrast, training the standard CNN on an augmented version of the dataset (with 4 rotated versions of all the training galaxies, one for each rotation angle), would require double the time.

\begin{figure}
    \centering
    \includegraphics[width=0.5\linewidth]{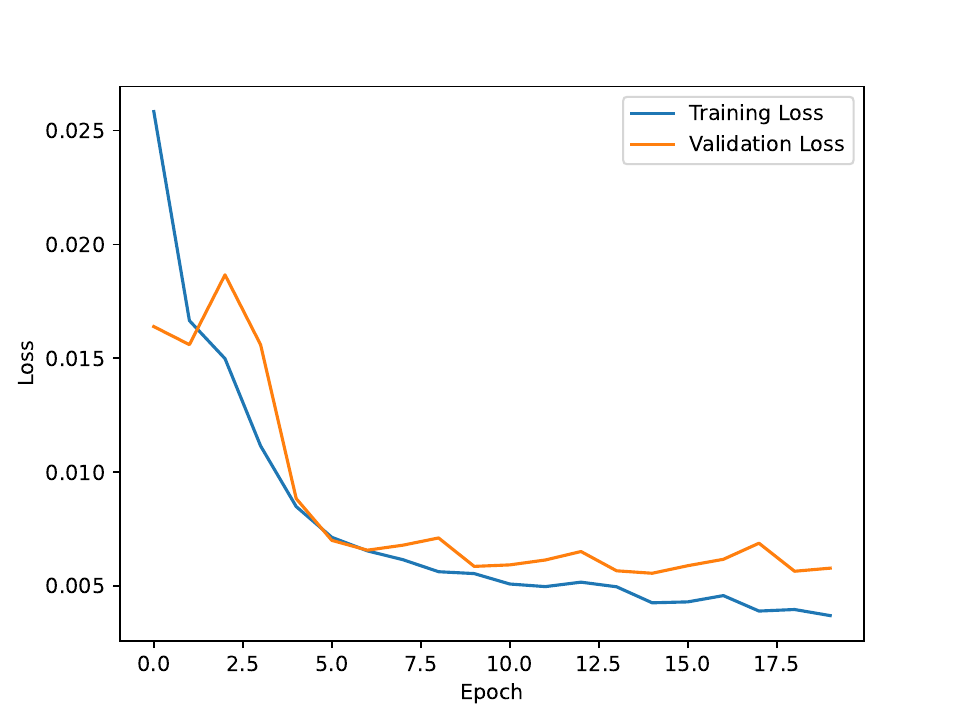}
    \caption{Training and validation loss behaviour for run 1 out 20 of the \ECNN instances.}
    \label{fig:loss}
\end{figure}

In \cref{fig:loss} we show the behaviour of the training and validation losses for an example run. It can be seen how the model starts learning (both the training and validation loss decrease) up to $\sim10$ epochs, where the validation loss flattens although the training loss continue decreasing. This is the typical behaviour when the model starts over-fitting. By using an early-stopping criteria we avoid entering the over-fitting region and keep the model with lower validation loss. We trace the origin of the over-fitting back to the reduced size of the training set (on average, about 1400 galaxies, with imbalanced classes, where spirals dominate) due to the bootstrapping. In particular, the minority class (ellipticals, which make up 23\% of the sample) will exhibit larger fluctuations and with repetitions potentially encouraging memorization due to the high capacity of our network, thus leading to early over-fitting.

\section{Results} 
\label{sec:results}

In this section, we present the main results of the paper. First, we assess the effectiveness of the \ECNN in reconstructing the non-parametric spherical dark matter profiles, represented by the dark matter mass enclosed at $20$ radial bins (\cref{fig:mse,fig:scatter}). In addition, we present a thorough interpretability analysis of the model, aiming to understand what the model is really learning, for example, which pixels are more important (shown in \cref{fig:masked-analysis}) or which channels contain the relevant information (shown in \cref{fig:maskingchannels}). 

\subsection{Dark matter mass profile reconstruction}

\begin{figure}
    \centering
    \includegraphics[width=0.99\linewidth]{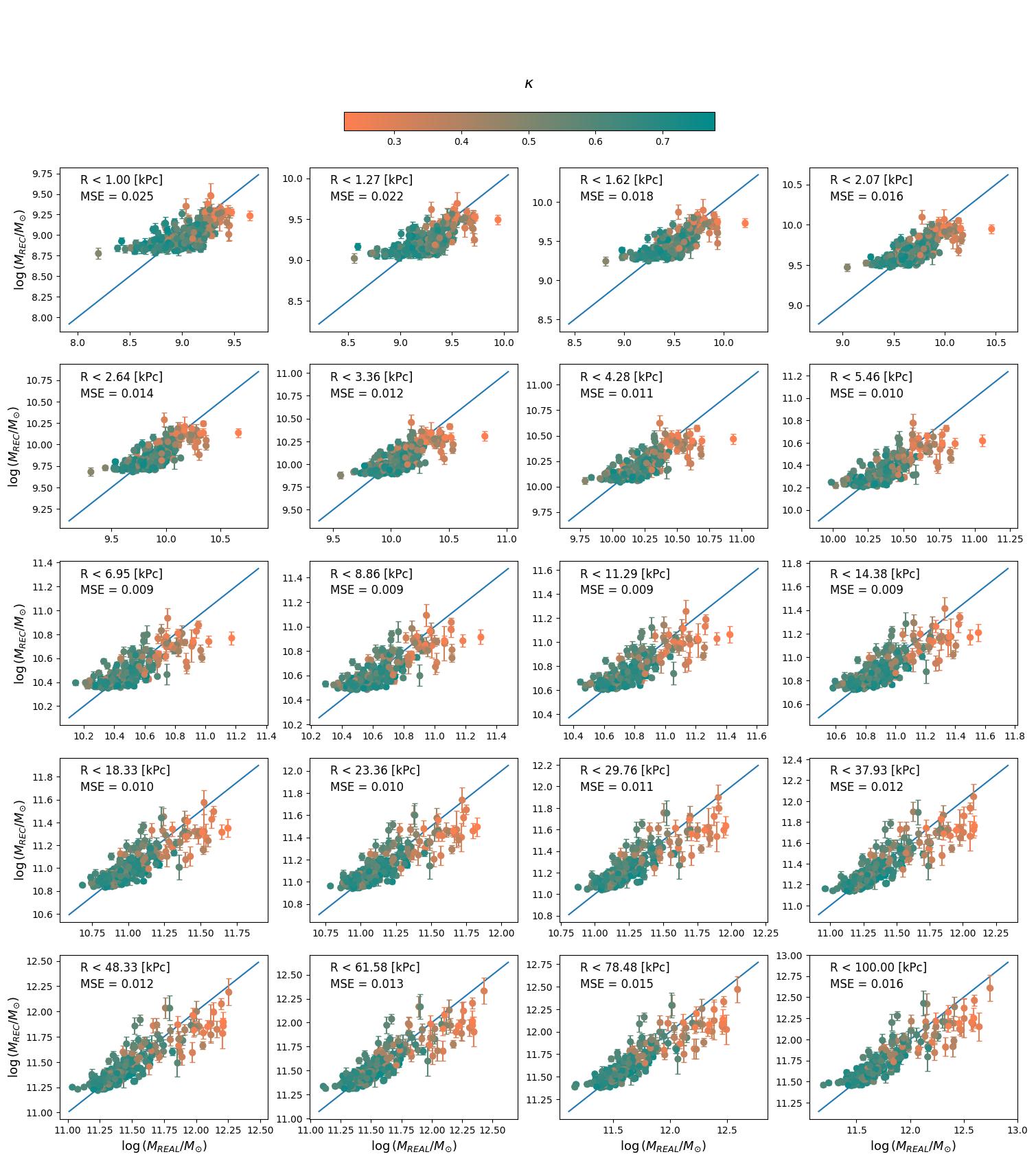}
    \caption{Scatter plot of the logarithm of the \ECNN estimated mass vs the logarithm of the real mass. Each panel correspond to the mass enclosed at a given radius as specified in the upper left corner. Each dot correspond to a test galaxy and its color represent the $\kappa$ value. The error bars correspond to the standard deviation of the $20$ estimated profiles.}
    \label{fig:scatter}
\end{figure}

\begin{figure}
    \centering
    \includegraphics[width=0.49\linewidth]{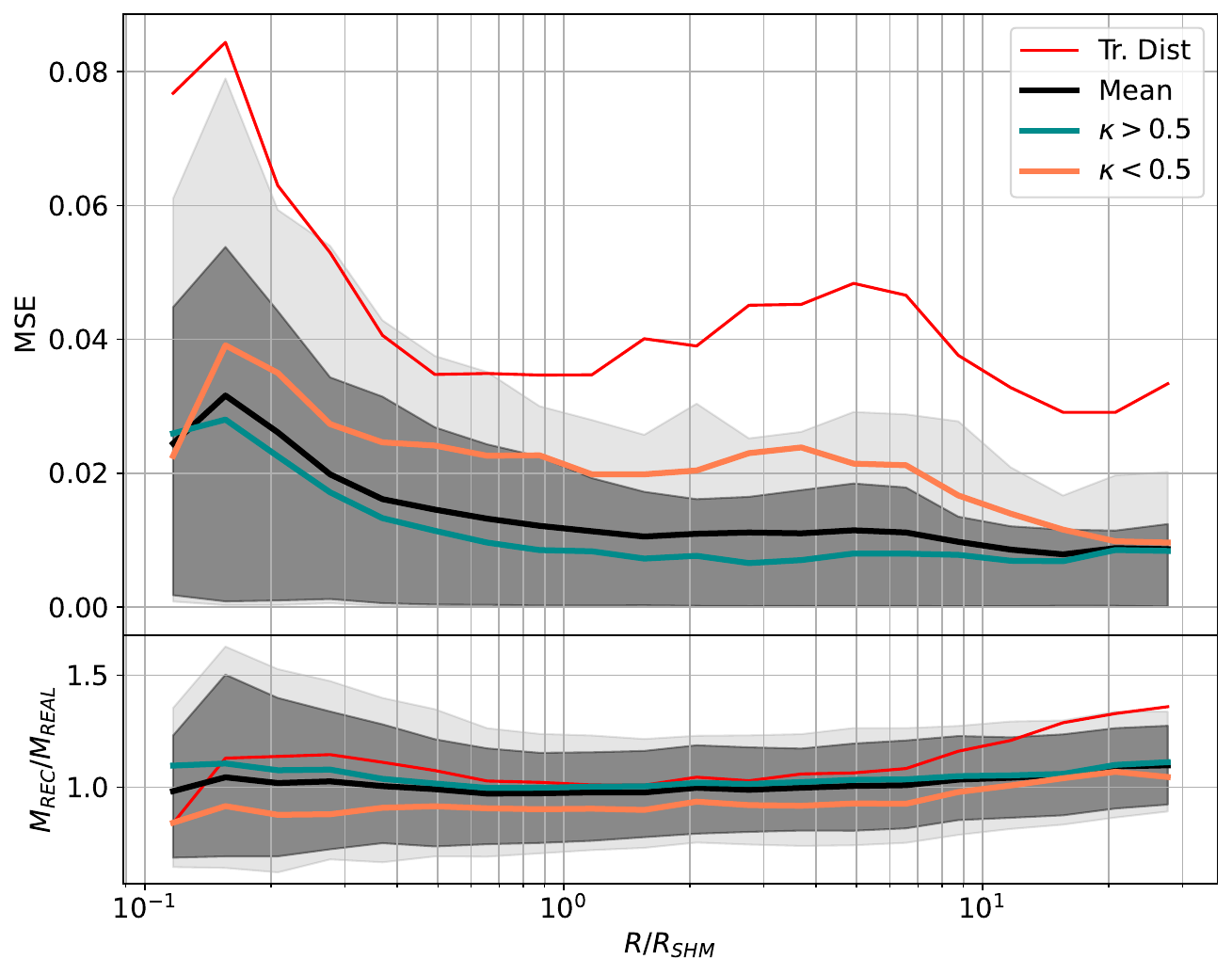}
    \includegraphics[width=0.49\linewidth]{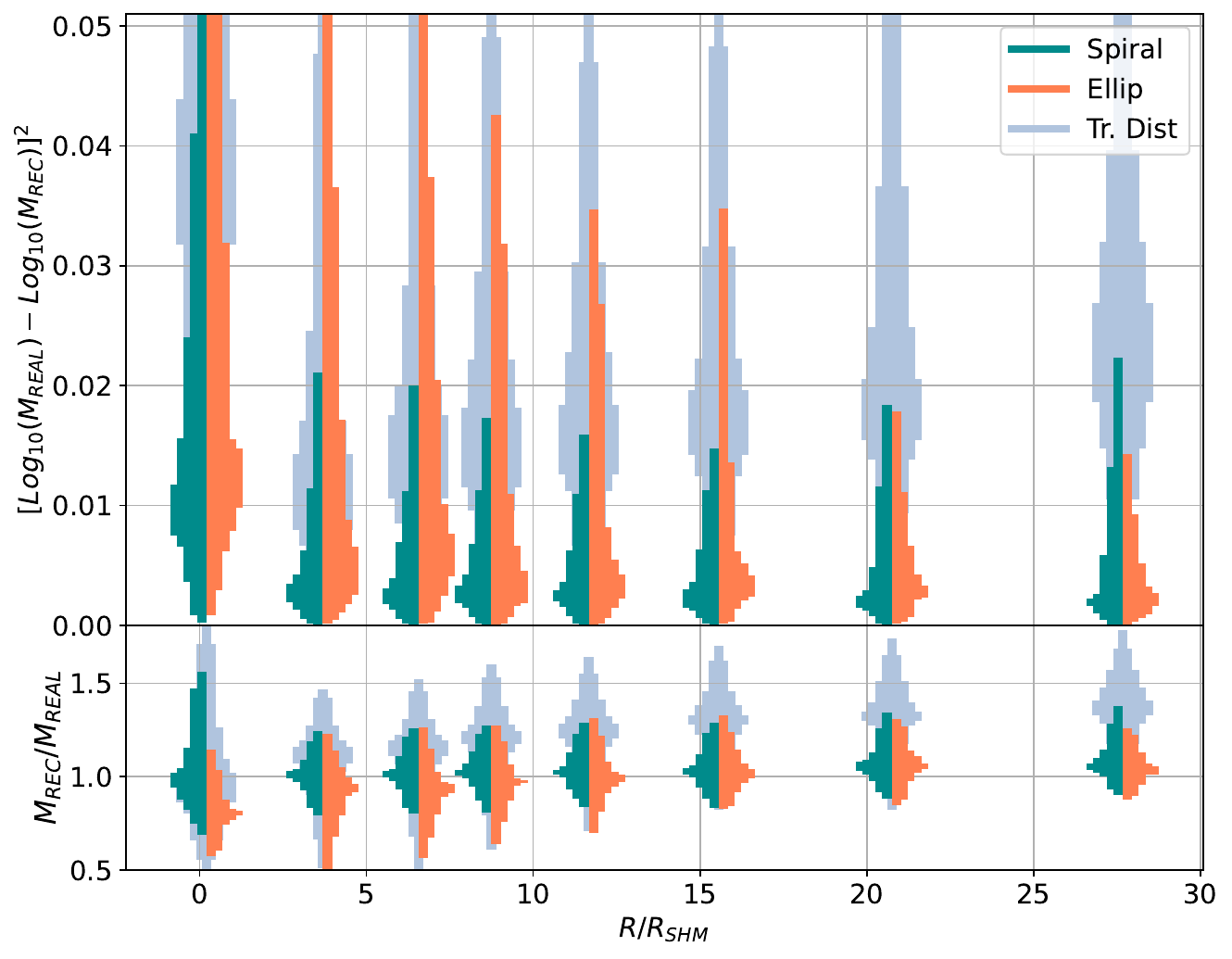}
    
    \caption{MSE of the enclosed mass at different radii normalized by the stellar half mass. 
    \textit{Left Panel:} In the upper panel we show the mean squared error per radial bin (black solid line) while in the lower panel we show the mean mass ratio (black solid line). In addition, we show the results when splitting the test set by the value of $\kappa$ (green and orange lines correspond to spiral and elliptical galaxies, respectively). For comparison, we add the mean of the training distribution as a solid red line.
     The gray areas correspond to $1$ and $2$ standard deviations.
     \textit{Right panel:} Violin plots of the squared error (upper panel) and the mass ratio (lower panel) in selected bins.
    In green (orange) we show the results for spiral (elliptical) galaxies and, for a comparison, we add the training distribution in gray. }  
    \label{fig:3normalized}
\end{figure}

The scatter plot in \cref{fig:scatter} shows the average of the reconstructed DM mass from the 20 \ECNN instances vs the true DM mass for all the galaxies in the test set. Each panel refers to the dark matter mass enclosed within a given radius, indicated in the left corner. 
In addition, we colour-code each galaxy according to its value of $\kappa$. A diagonal blue line shows what would be a perfect reconstruction, while in the upper-left part of each panel we indicate the MSE of the estimations at each radial bin. 
It can be seen that the model has a very good performance approaching the diagonal line in all the radial bins, but the first 4. 
This trend is also noticeable in the left panel of \cref{fig:3normalized}, where we show the MSE for all the radial bins normalized to the stellar half-mass radii $R_{SHM}$ of each galaxy, which enables a more physical comparison between the different galaxies.
This behaviour can be explained as in the inner part of galaxies the stellar component dominates over the dark matter one, making it more difficult to reconstruct the DM mass, in which case the model simply reverts to guessing the mean of the training distribution, as one might expect.

We investigate the performance of our model in different types of galaxies, classified using the ratio between rotational energy and total energy $\kappa$ defined in \cref{eq:kappa}.
We split the test set in spiral ($\kappa > 0.5$) and elliptical ($\kappa < 0.5$) galaxies and compute the MSE in each radial bins for both classes.
The result for spiral and elliptical galaxies are shown in green and orange respectively in \cref{fig:3normalized}, where we show the distributions of the MSE (top-right panel), and the ratio of reconstructed and real DM mass (bottom-right panel).
We show here only $8$ bins for visualization reasons, without altering the conclusions from the full binning analysis.
Our method performs equally well in capturing the total mass of both elliptical and spiral galaxies in our sample,  with only a slight tendency ($\lesssim$10\%) to overestimate.
Furthermore, the method performs slightly better for spiral galaxies in the innermost regions. 
It is interesting to note that the mass reconstruction achieves its best performance at a radius of $2 \times R_{SHM}$ (the stellar half-mass radius), beyond which the improvement flattens out. This behaviour can be interpreted as the region where the information content is maximal, including significant contributions from both interferometry and photometry.
In addition, it can be seen that the performance worsens towards the tail of the mass distributions. This is a well-known behaviour in machine learning models, since algorithms start entering the extrapolation region in the vicinity of the boundary of the training set~\cite{dakhmouche25}. In our case, the trend is especially pronounced toward the low-mass end of our galaxy set, where the model overestimates the mass.

It is worth highlighting that the dark matter profiles of the training galaxies only span a limited range ($[10^{11}-10^{13} M_{\odot}]$ of total dark matter mass).
This range can be considered as prior information about the dark matter profiles before any analysis and depends on the characteristic of the cosmological simulation used (box size, dark matter and baryonic resolution, hydrodynamical code, initial conditions,  etc.). 
The mean profile of the training galaxies (i.e. the mean enclosed mass at each radial bin computed over the galaxies in the training set) can thus be used as a first guess to the dark matter profile of the testing galaxies before any image is provided or analyzed.
Taking this into account, we compute the MSE and the mass-ratio obtained when assigning to each galaxy of the test set the mean dark matter profile of the training set. 
We call this the ``training distribution'' throughout our paper. The MSE of the training distribution -- which we call the `training MSE' -- can be regarded as a baseline against which to quantify the improvement obtained with our model. 
The training distribution results are shown by red lines in the left panels and in gray histogram in the right panel of  \cref{fig:3normalized}. 
It can be seen how the reconstructed mass distributions from our model has an average value (across all bins) of $0.013$, as opposed to the training MSE, which is $0.044$. 
This is a demonstration that the \ECNN is actually learning and not simply memorizing the training distribution.
It also can be seen that our model is slightly overestimating the dark matter profiles of spiral galaxies while underestimating the one of elliptical galaxies.

\begin{figure}[h]
    \centering
    \includegraphics[width=1\linewidth, trim={0 2cm 0 4cm}, clip]{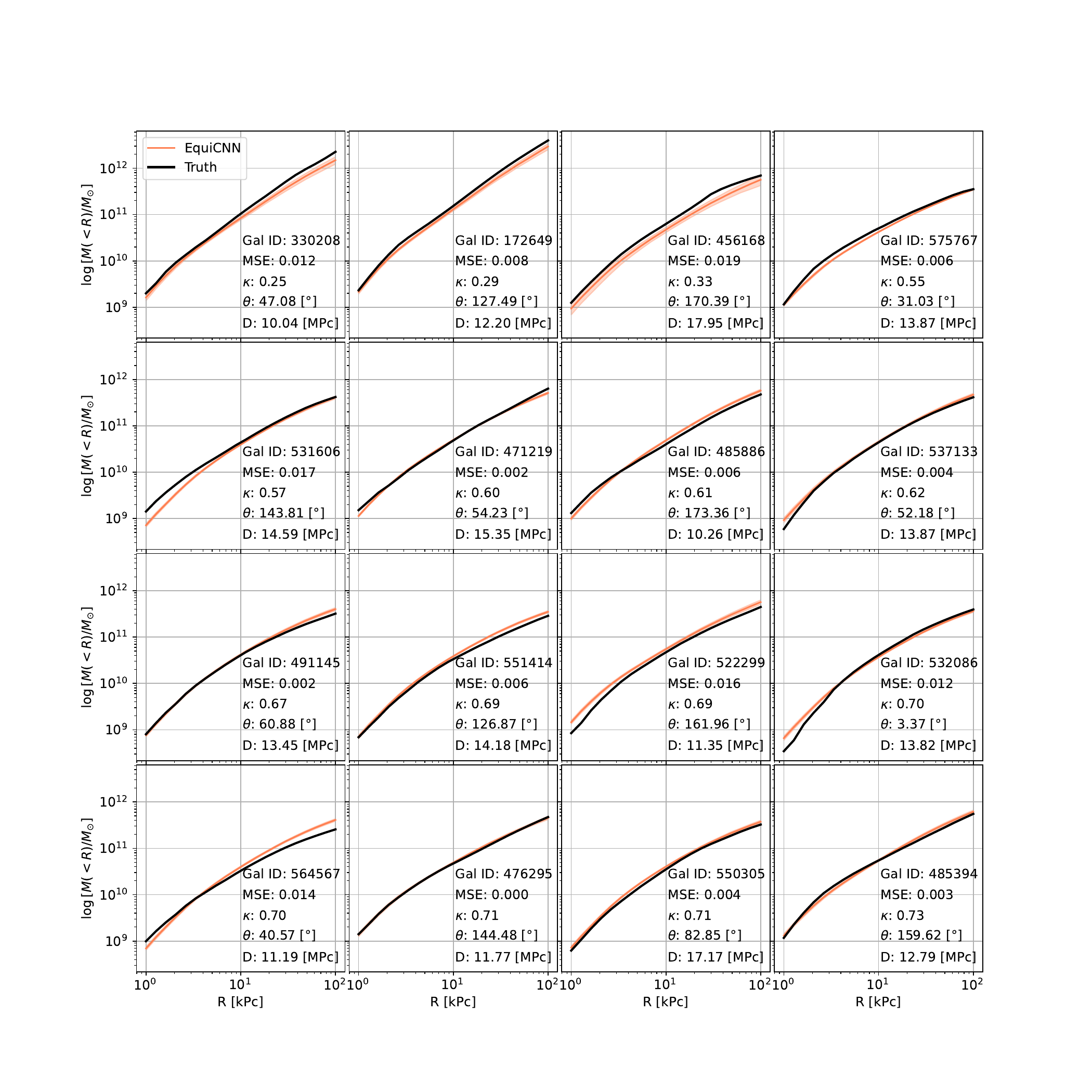}
    \caption{Average DM profile reconstructed with the 20 \ECNN instances (orange, with shaded region giving the standard deviation among the 20 instances), in comparison with the actual DM profile (black) for the simulated galaxies shown in Figs.~\ref{fig:sdss_examples} and ~\ref{fig:h1_examples}. In each panel we indicate some properties of the galaxy being analysed, including its MSE averaged over all radial bins. For a reference we add, in the bottom right corner, the corresponding subhalo ID, the fraction of rotational energy $\kappa$ and the distance $D$ and inclination angle $\theta$ of the rendered configuration.}
    \label{fig:individualProfiles}
\end{figure}

We show in \cref{fig:individualProfiles} the reconstructed profiles of the same galaxies of \cref{fig:sdss_examples,fig:h1_examples}. We observe no systematic bias and a reconstruction that follows closely the ground truth.
In addition, in \cref{sec:normalized} we show the results in physical radial bins without any normalization.

\subsection{Masked analysis}

In this section, we analyse the results, in order to understand what physical information the models are actually using to estimate the dark matter profile.
For this, we analysed the reconstructed profiles after masking different channels and different regions inside the channels, aiming to determine the inputs that have a higher impact on the final results.

\subsubsection{Masking radial bins}

\begin{figure}
    \centering
    \includegraphics[width=0.47\linewidth]{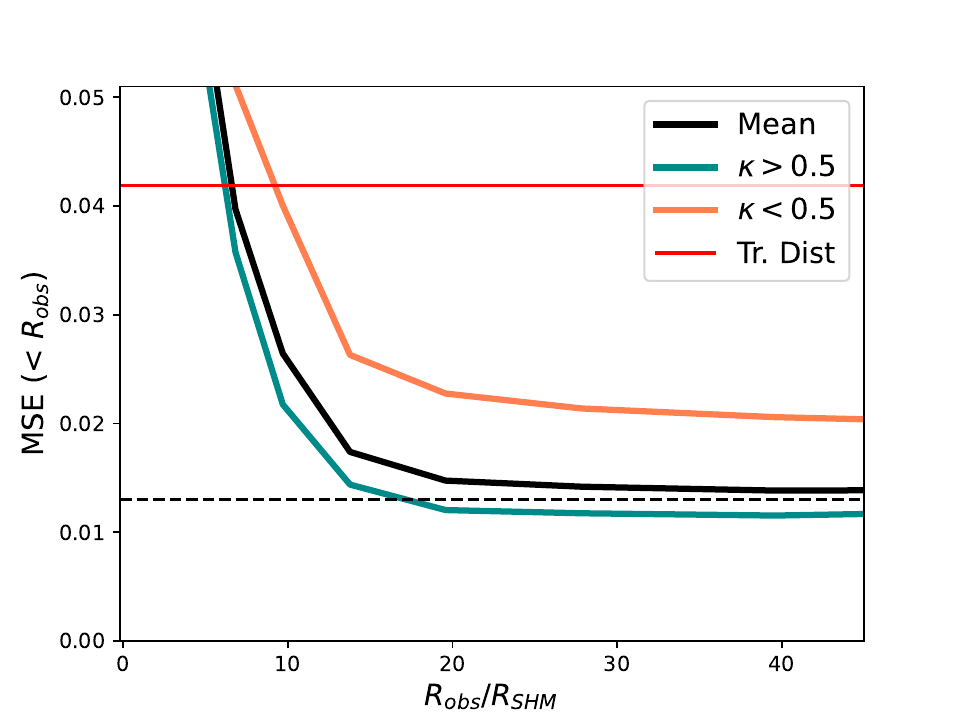}
    \caption{MSE as a function of the observed radii $R_{obs}$ normalized to $R_{SHM}$. In black solid line we depict the results for the full test set, while in green and orange we show the results of the spiral and elliptical galaxies respectively. 
    For a clearer comparison we add the mean of the training distribution (red horizontal line) and the MSE obtained with the full images (black dashed line).}
    \label{fig:masked-analysis}
\end{figure}

As a first test, we mask, for each galaxy, all the pixels outside a given radius ($R_{obs}$), and estimate the corresponding dark matter profiles with the masked inputs.
In \cref{fig:masked-analysis} we show the MSE obtained when masking outside regions, i.e. the images only contain information up to a given radius $R_{obs}$.
The fidelity of the reconstruction in the outer parts of the galaxies decreases when those are masked out. 
In particular, it can be seen that, when masking almost the whole image (i.e., for small values of $R_{obs}$), the reconstructed profiles give worst results that the one obtained when using the mean of the training distribution (red horizontal line), i.e, the model is not improving over what we can estimate just from the training set.
Also, it can be noticed that, when increasing $R_{obs}$, the results stabilize at the MSE obtained with the full images (black dashed line).
These results give confidence that the model is using the inner pixels, where the galaxy is located, to estimate the dark matter profiles, which is expected from physical grounds.

\subsubsection{Masking channels}

As a second test, we estimate the dark matter profiles after masking individual channels, in order to ascertain where the information comes from.

\begin{figure}
    \centering
    \includegraphics[width=0.48\linewidth]{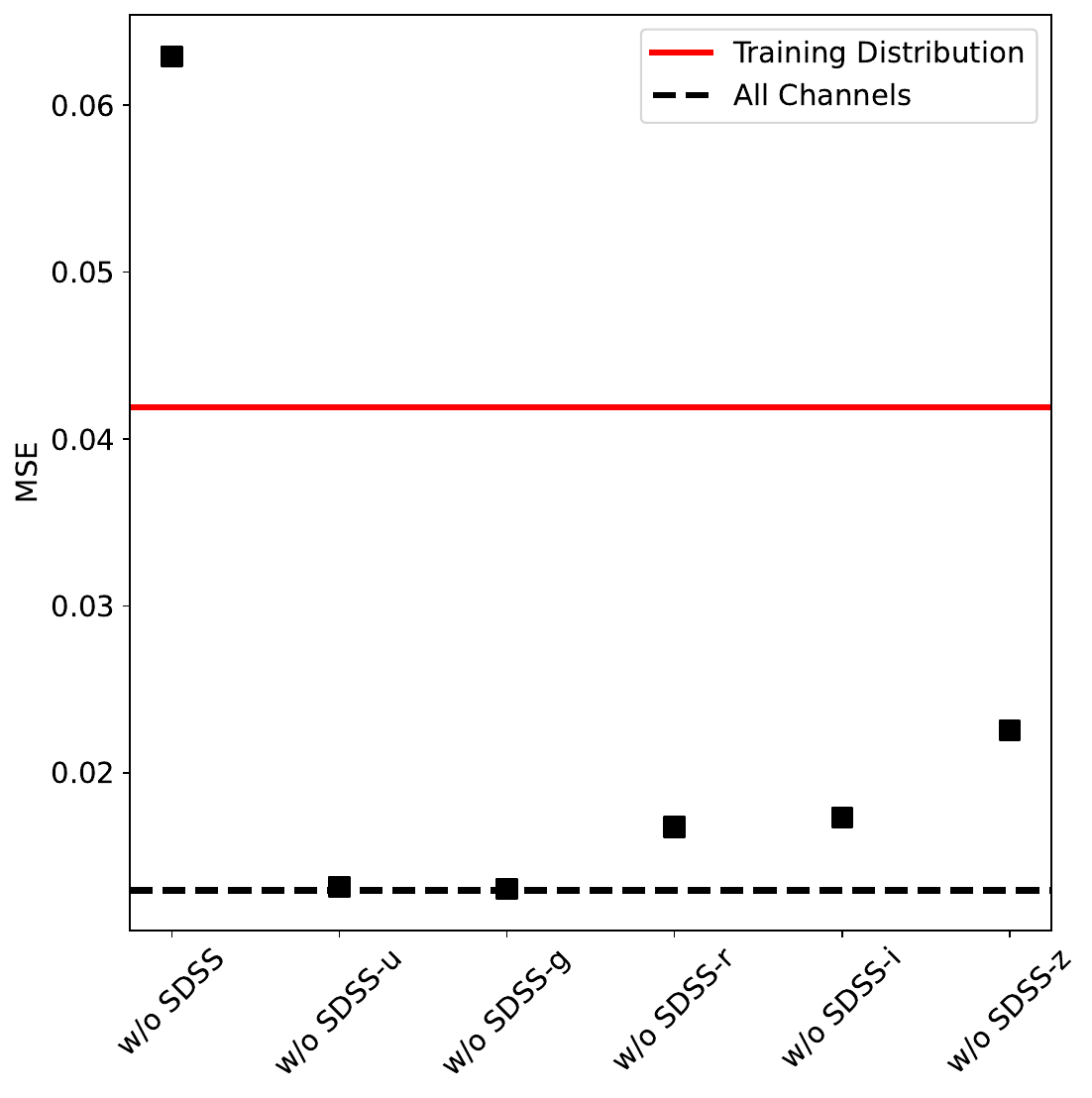}
    \includegraphics[width=0.48\linewidth]{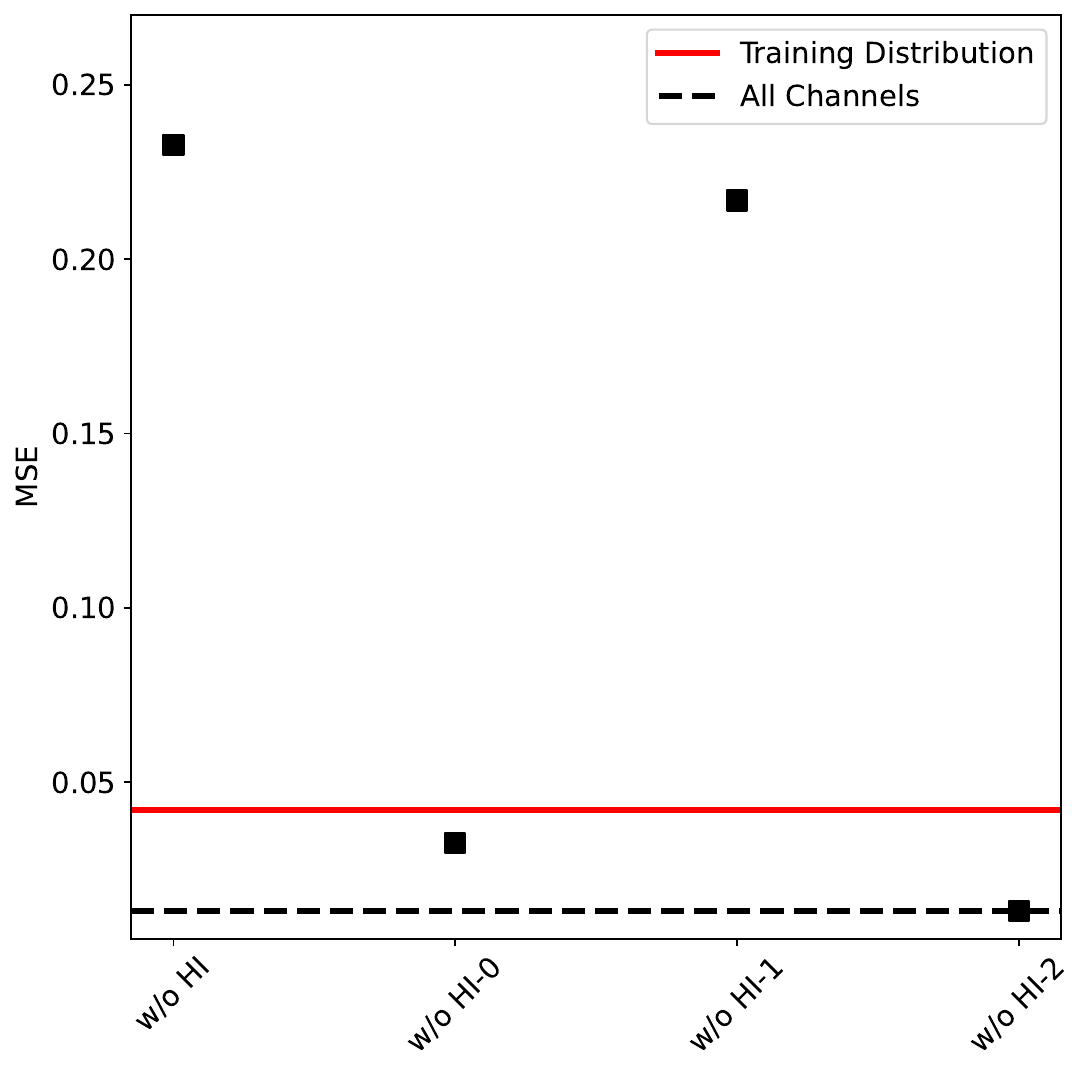}
    \caption{MSE obtained when masking different channels (squares), compared with the MSE obtained when using all data channels (black horizontal line) and that of the training sample (red horizontal line). }
    \label{fig:maskingchannels}
\end{figure}

In the left panel of \cref{fig:maskingchannels} we show the MSE obtained when masking the photometric channels while in the right panel we show the results obtained when masking the HI channels.
We notice that when removing entirely one type of information: either the full SDSS photometry --which incapsulates all information about the stellar mass-- or the full HI spectroscopic data --which includes information about the gas dynamics-- the performance worsens dramatically with respect to the MSE that one would have obtained by simply using the average from the training set (indicated by the red line), which however does incorporate both kinds of data. This is therefore understandable on physical grounds, and confirms that the models are processing the information in an interpretable way.

It is also worth noting that masking the SDSS $u$ or $g$ bands has a negligible impact on the results, whereas masking the $r$, $i$, or $z$ bands leads to a deterioration in performance.
Furthermore, as expected, masking the map of the first momentum of HI emission (HI$-$1) worsen the performance producing worse results than the training set average, which in contrast does contain such information. The interpretation of this is that this map encodes the information about the velocity of the gas, and removing it equates to removing the entire content of dynamical information, therefore strongly degrading performance, as observed.

\section{Discussion} \label{sec:discussion}

Machine learning tools have become indispensable for analysing astronomical data.
Thanks to their great flexibility these methods have surpassed traditional techniques in different domains.
Despite much promise, however, results produced by machine learning methods can be hard to interpret and may hide biases and undetected model miss-specifications.
In this work, we trained equivariant convolutional neural networks to estimate the dark matter profiles from photometric images and HI-emission maps in a controlled, ``{\tt in silico}'' environment, as a first step to assess the viability and performance of this approach, improving on similar work in the past.
We also carried out an interpretability analysis, highlighting the channels and regions that have the strongest impact on the final results.

Our models were trained on simulated galaxies with total dark matter masses ranging from $[10^{11}-10^{13} M_{\odot}]$. This mass range serves as baseline information about dark matter and allows us to quantify the improvements achieved by our model. As the main result, we demonstrate and quantify the quality of the dark matter profile reconstruction for galaxies within our {\it in silico} environment. In particular, the distributions reconstructed by the model after training show substantial improvement compared to the baseline profiles, providing clear evidence that the model is genuinely learning rather than memorizing the training set. 
Specifically, the MSE averaged over all bins decreases from $0.044$ from the training distribution to $0.013$ with the reconstructed profiles. These results can be contrasted with the model of \cite{Delosrios2111}, which used a standard convolutional neural network architecture and a training set 2/3 the size of the one adopted here. The square root of our MSE on the validation set achieves a value of 0.11, which can be directly compared with the value of $\sim 0.20$ reported in \cite{Delosrios2111}, thus demonstrating an improvement of almost a factor of 2. It is also worth to mention here a comparison of our work with the results of previous analyses that have applied machine learning methods to estimate the mass of galaxies. In \cite{Villanueva21} the authors infer the halo mass given the positions, velocities, stellar masses, and radii of the galaxies it hosts achieving a $\sim0.2$ dex accuracy. In \cite{Sarrato25}, the authors estimate the dynamical mass profiles of dispersion-supported galaxies using stellar data, and the MSE achieved is of order $\sim 0.01$, similar to the one obtained here.

The method accurately captures the total mass ($\lesssim 10\%$) for both spiral and elliptical galaxies. It performs better at reconstructing the innermost regions of spiral ($\kappa>0.5$) than that of ($\kappa<0.5$) galaxies. The reconstruction reaches its optimal performance at a radius of $2 \times R_{SHM}$ (the stellar half-mass radius), beyond which the results remain essentially constant. This reflects the region where the information content provided by the observable data is maximized.

The interpretability analysis demonstrates that masking the central region of the galaxy leads to poorer reconstruction results, whereas masking outer regions beyond the galaxy has no significant effect. Furthermore, masking different channels highlights the importance of each data type: excluding photometric information degrades performance beyond the training distribution, and omitting the HI first-moment map, which traces gas velocity, similarly reduces the quality of the reconstruction.

These results are promising, yet several steps remain before applying this method to real galaxy data. In this study, we trained our models using the TNG-100 simulation \cite{Nelson2018,Pillepich2018}. Other simulations, such as e.g.~EAGLE \cite{eagle}, SIMBA \cite{simba}, or Magneticum \cite{magneticum}, may yield different results on small scales depending on the simulation suite and implemented subgrid physics (as shown for example in~\cite{Contardo}). A robust machine learning model should generalize across different simulation frameworks, accurately reconstructing mass profiles regardless of subgrid physics implementation.

Additionally, creating synthetic images requires fixing several parameters, such as spectral energy distribution of stars, dust composition, dust-to-metal ratio, etc., and it is not always clear which choices are most realistic: for example, whether one should adopt the population synthesis model of Ref.~\cite{Bruzual2003} or a fixed dust-to-metal ratio (see the discussion in \cite{Huertas2019} for further details). Variations in noise modeling, point spread function, and resolution can significantly affect downstream analyses, making robustness essential. Observational features, including sky background, pixel number, resolution, field of view, and bright stars, vary widely across surveys and can strongly impact the results; thus, the model must accommodate this variability \cite{Dominguez19}. Fortunately, in view of large future surveys like LSST, there exist sophisticated and large simulations of telescope output, which can be adapted to investigate the above issue.

In this work, we present a proof of concept for an equivariant network capable of reconstructing dark matter profiles, leaving broader analyses and application to real galaxies for future work.

\section{Conclusions} \label{sec:conclusion}

In the work presented in this paper, we have applied machine learning methods to estimate the dark matter profile of galaxies in the controlled, {\tt in silico} environment of numerical cosmological simulations. While the analysis of their performance shows very promising results,  we believe that several additional steps are required before the method proposed can be reliably applied to real world observations. The robustness of the generalization performance to model misspecification and domain shift between train and test data needs to be thoroughly assessed, for example, by training on one simulation but deploying on another suite with different implementations of subgrid physics.
The current architecture requires images of fixed size, something that we plan to improve upon to allow for variable size input, as this is important when dealing with real observations. Furthermore, observational data coming from different instruments will have different instrumental characteristics, like resolution, PSF and noise, which need to be faithfully simulated in the training set. The upcoming, large datasets from the Vera Rubin telescope will no doubt provide a rich testing ground for these techniques, with the promise of improving manifold our understanding of the dark matter distribution in galaxies.

\section*{Acknowledgements}

We thank an anonymous referee for useful suggestions. RT acknowledges co-funding from Next Generation EU, in the context of the National Recovery and Resilience Plan, Investment PE1 Project FAIR ``Future Artificial Intelligence Research''. This resource was co-financed by the Next Generation EU [DM 1555 del 11.10.22]. RT is partially supported by the Fondazione ICSC, Spoke 3 ``Astrophysics and Cosmos Observations'', Piano Nazionale di Ripresa e Resilienza Project ID CN00000013 ``Italian Research Center on High-Performance Computing, Big Data and Quantum Computing'' funded by MUR Missione 4 Componente 2 Investimento 1.4: Potenziamento strutture di ricerca e creazione di ``campioni nazionali di R\&S (M4C2-19)'' - Next Generation EU (NGEU). 
FI is also supported by the research grant number 2022E2J4RK ``PANTHEON: Perspectives in Astroparticle and Neutrino THEory with Old and New messengers'' under the program PRIN 2022 funded by the Italian Ministero dell’Università e della Ricerca (MUR).

\newpage
\section*{Appendix}
\appendix

\section{Configuration details of SKIRT and MARTINI} \label{sec:skirtmartini}

Following the procedure described in~\cite{RodriguezGomez2019}, for each simulated galaxy, we create synthetic images that replicate the $5$ SDSS photometric bands~\footnote{\href{https://www.sdss.org/dr16/imaging/imaging_basics/}{\url{https://www.sdss.org/dr16/imaging/imaging_basics/}}}.

To this end, the spectral energy distribution (SED) of each star particle is computed using the population synthesis model of Bruzual \& Charlot~\citep{Bruzual2003}.
This model requires, as input, the initial mass, metallicity and age of each star particle, all of which can be obtained from the simulation. Each SED is sampled using $1000$ logarithmic wavelength bins in the range between $0.09$ to $100$ $\mu \textrm{m}$.
Similarly to \cite{RodriguezGomez2019} for dust modelling, we adopt the multicomponent dust mixture proposed by \cite{Zubko2004} and assume that it follows the distribution of star-forming gas with a constant dust-to-metal mass ratio of $0.3$. In addition, we model the dust emissivity as a modified blackbody spectrum  and track it over $1000$ logarithmic bins spanned between $0.09$ and $100$ $\mu\textrm{m}$.

To produce the emission data-cubes, we run the Monte-Carlo radiative transfer simulations with $10^7$ photon packets, using a frame instrument with $1000$ logarithmically spaced bins in the range between $0.01$ and $3.7$ $\mu\textrm{m}$. The resulting datacubes are then convolved with the five SDSS \textit{ugriz} broadband filters to obtain the corresponding photometrical images.
Finally, to produce realistic mock observations, we incorporate the sky background contribution by adding a random Gaussian noise to each image pixel, where the appropriate mean and variance values for each band were taken from SDSS DR16~\footnote{\href{https://www.sdss.org/dr16/imaging/other_info/}{\url{https://www.sdss.org/dr16/imaging/other_info/}}}.
We further convolve each image with a Gaussian kernel filter with a standard deviation that matches the width of the point spread function (PSF) in each of the SDSS wavebands. The adopted PSF and median sky brightness values are contained in \cref{tab:obs_corrections}.

\begin{table}[H]
	\centering
	\begin{tabular}{c|c|c}
		Filter & Seeing PSF [arcsec] & Sky brightness [mag/arcsec$^2$] \\ \hline
		$u$ & 1.53 & 22.01 \\
		$g$ & 1.44 & 21.84 \\
		$r$ & 1.32 & 20.84 \\
		$i$ & 1.26 & 20.16 \\
		$z$ & 1.29 & 18.96
	\end{tabular}
	\caption{Seeing PSF and median sky brightness used in the creation of SDSS images. The values are adopted from SDSS DR16.}
	\label{tab:obs_corrections}
\end{table}

In parallel, for each simulated galaxy, we also generate a mock HI observation following the methodology described in~\cite{martini} and implemented in the \texttt{MARTINI} code~\footnote{\href{https://github.com/kyleaoman/martini}{https://github.com/kyleaoman/martini}}.
We use a \texttt{MARTINI} configuration to resemble data from the THINGS survey \cite{things}, with $64$ channels with spectral resolution of $5 \, \textrm{km/s}$, a cubic spline kernel for particle smoothing, and a Gaussian PSF with a full width at half maximum (FWHM) of $10''$, truncated at $3 \times$FWHM. A Gaussian noise model with a root-mean-square value of $5 \times 10^{-6} \, \textrm{Jy/arcsec}^{2}$ is also applied.
Finally, from the HI emission data-cubes we computed the corresponding line-of-sight intensity, average velociwty and velocity dispersion maps, which can be obtained by evaluating the 0th, 1st and 2nd moments of the emission intensity across the available frequency channels. In addition, as suggested in~\cite{martini}, we masked regions where the HI intensity dropped below $0.2 \, \textrm{Jy/beam}$ (equivalent to an HI column density of $10^{19.5} \, \textrm{atoms} \, \textrm{cm}^{-2}$)
This resulted in the final images, which were later used for training the neural network. 

\section{Results as a function of physical radial bins}\label{sec:normalized}

\begin{figure}[H]
    \centering
    
    \includegraphics[width=0.49\linewidth]{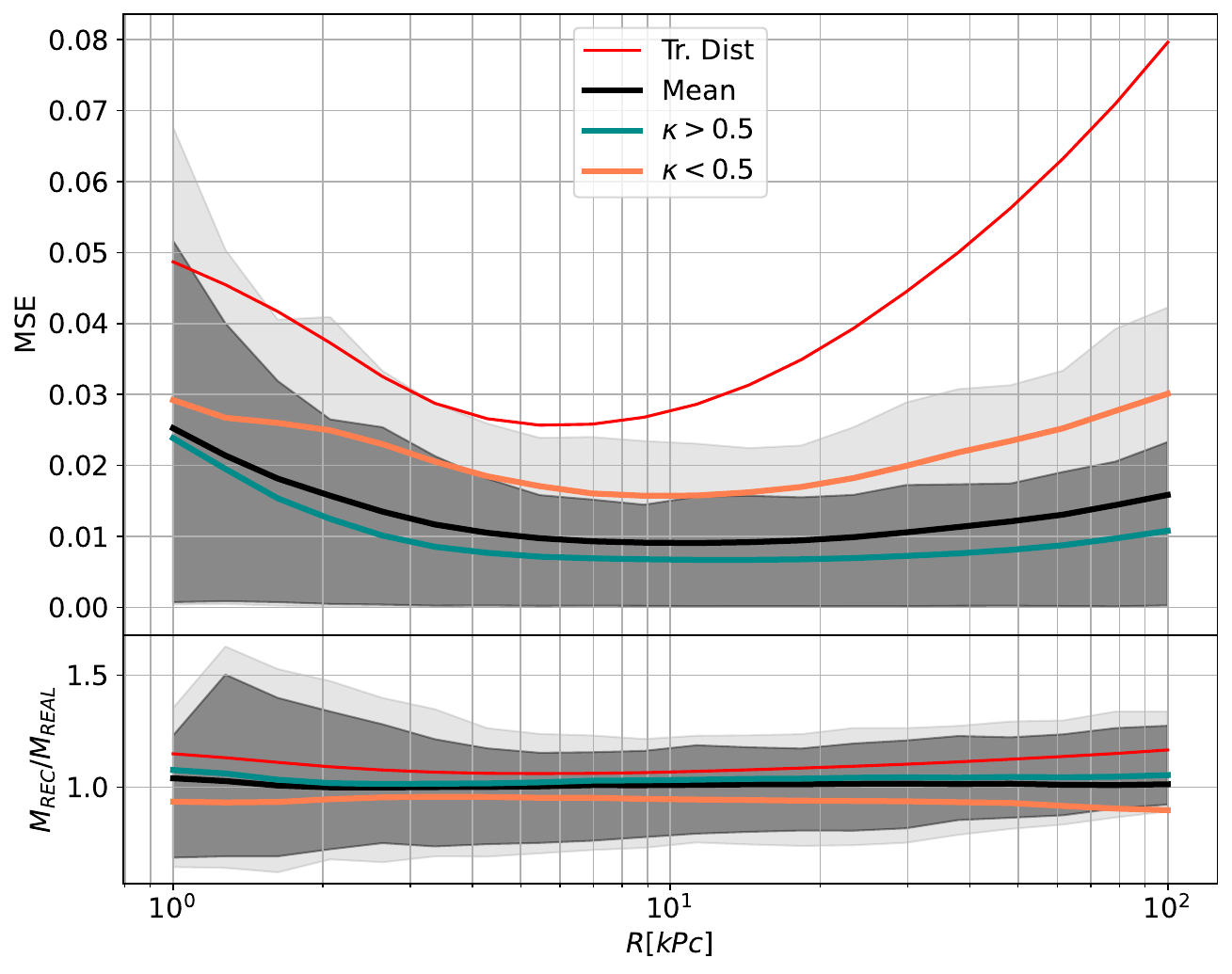}
    \includegraphics[width=0.49\linewidth]{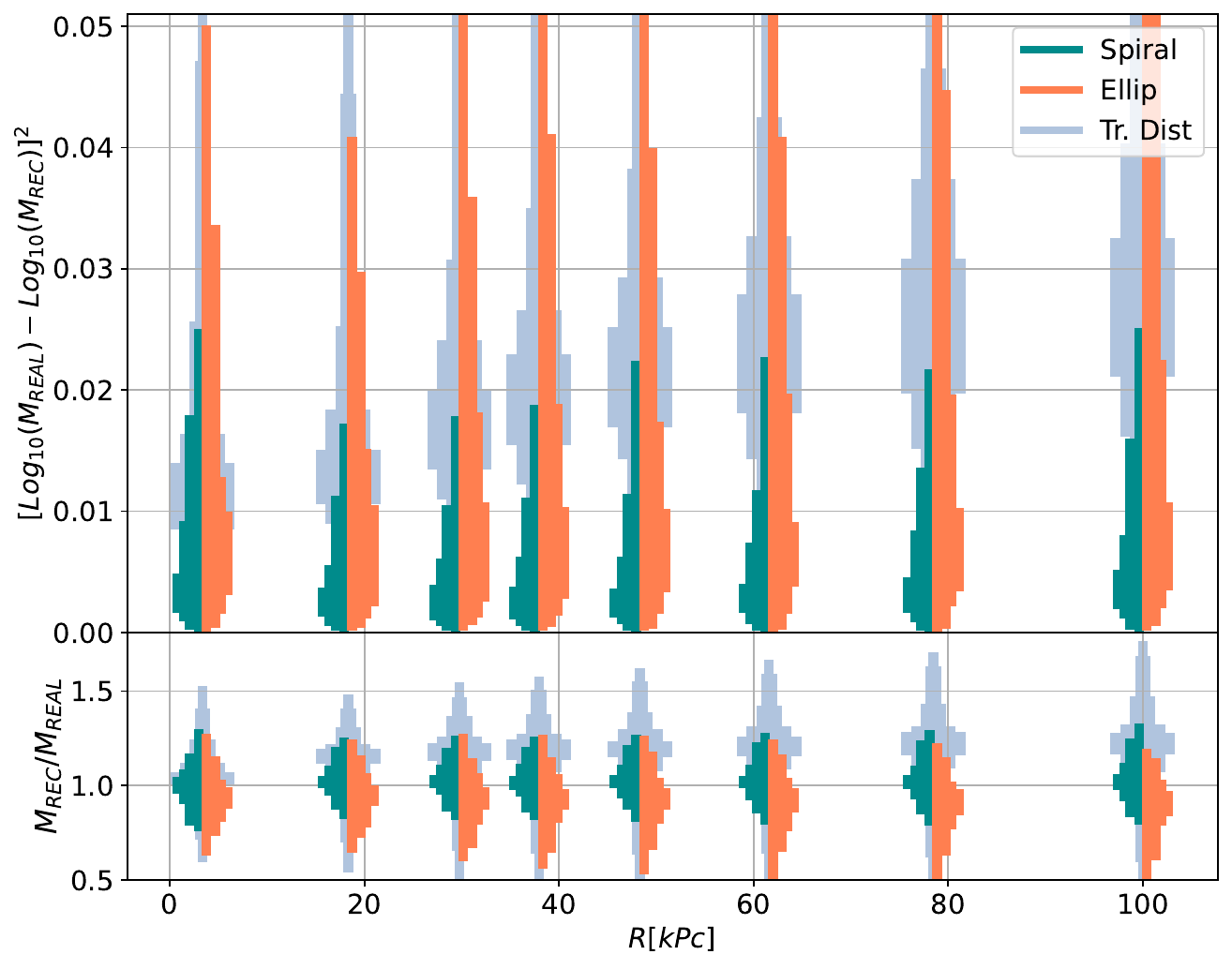}
    
    \caption{MSE of the enclosed mass at different physical radii.
    \textit{Left Panel:} In the upper panel we show the mean squared error per radial bin (black solid line) while in the lower panel we show the mean mass ratio (black solid line). In addition, we show the results when splitting the testset taking into account $\kappa$ (green and orange lines correspond to spiral and elliptical galaxies respectively). For a better comparison, we add the mean of the training distribution as a solid red line.
     The gray areas correspond to $1$ and $2$ standard deviations from the estimated mean.
     \textit{Right panel:} Distribution of the squared error (upper panel) and the mass ratio (lower panel) in selected bins.
    In green (orange) we show the results for spiral (elliptical) galaxies and, for a better comparison, we add the training distribution in gray. }
    \label{fig:mse}
\end{figure}

For completion, we show in \cref{fig:mse} the mean squared error and mass ratio in physical radial bins without any normalization.

\newpage

\section{Comparing \ECNN\ and a standard CNN on rotated galaxies} \label{app:rotation-test}

\begin{figure}[h]
\centering
\begin{tikzpicture}[
  font=\footnotesize,
  node distance=4mm,
  >={Stealth[length=2mm]},
  obox/.style={draw, rounded corners, inner sep=1pt},
  ibox/.style={draw, rounded corners, align=center, minimum height=5.5mm, text width=20mm},
  title/.style={font=\bfseries}
]

\tikzset{ 
   obox/.style={ draw, rounded corners=4pt, very thick, inner sep=2mm, fill=gray!3 
   }, 
   ibox/.style={ draw, rounded corners=2pt, thick, inner xsep=2mm, inner ysep=1.5mm, align=center, font=\footnotesize, fill=white}, 
   title/.style={font=\bfseries}, 
   note/.style={font=\scriptsize, align=left} 
   
} 

\newcommand{\fcPreviewCount}{4} 
\newcommand{\fcPreviewShiftX}{+2mm} 
\newcommand{\fcPreviewShiftY}{+3mm} 
\newcommand{\fcPreviewDraw}{black!35} 
\newcommand{\fcPreviewFill}{gray!3} 

\node[ibox, text width=16mm] (inp) {Input\\$(C,H,W)$\\};

\newcommand{\RtwoBlock}[4]{%
  \begin{pgfonlayer}{main}
    \node[ibox, right=9mm of #1] (relu-#4) {ReLU };
    \node[ibox, above=1mm of relu-#4] (bn-#4) {InnerBN };
    \node[ibox, above=1mm of bn-#4] (conv-#4)  {2DConv};
    \node[ibox, below=1mm of relu-#4] (drop-#4) {Dropout2d };
    \node[ibox, below=1mm of drop-#4, text width=14mm] (pool-#4) {MaxPool $2{\times}2$\\ };
  \end{pgfonlayer}
  \begin{pgfonlayer}{frame}
    \node[obox, fit=(conv-#4)(pool-#4),text width=22mm, label={[title]above:#2}] (#4) {};
  \end{pgfonlayer}
  
  \begin{pgfonlayer}{background} 
    \foreach \i in {2,1} {%
    \draw[rounded corners=4pt, very thick, draw=\fcPreviewDraw, fill=\fcPreviewFill] 
    ($(#4.north east)-(\i*\fcPreviewShiftX,\i*\fcPreviewShiftY)$) rectangle 
    ($(#4.south west)-(\i*\fcPreviewShiftX,\i*\fcPreviewShiftY)$); }%
  \end{pgfonlayer} 
}

\RtwoBlock{inp}{3 $\times$ Conv. blocks}{Out: $5\times$ regular, $|G|{=}4$}{B1}

\draw[->] (inp.east) -- (B1.west);

\node[ibox, right=10mm of B1, text width=20mm] (gp) {GroupPooling\\};
\draw[->] (B1.east) -- (gp.west);

\node[ibox, below=6mm of gp, text width=15mm] (fl) {Flatten};
\draw[->] (gp.south) -- (fl.north);

\newcommand{\FCblock}[4]{%
  \begin{pgfonlayer}{main}
    \node[ibox, below=12mm of #1]  (fc-#4) {Fully-Connected layer};
    \node[ibox, below=1mm of fc-#4] (bn-#4) {InnerBN (opt)};
    \node[ibox, below=1mm of bn-#4] (relu-#4) {ReLU };
  \end{pgfonlayer}
  \begin{pgfonlayer}{frame}
     \node[obox, fit=(fc-#4)(relu-#4), label={[title]above:#2}] (#4) {};
  \end{pgfonlayer}

  \begin{pgfonlayer}{background} 
    \foreach \i in {4,3,2,1} {%
    \draw[rounded corners=4pt, very thick, draw=\fcPreviewDraw, fill=\fcPreviewFill] 
    ($(#4.north east)-(\i*\fcPreviewShiftX,\i*\fcPreviewShiftY)$) rectangle 
    ($(#4.south west)-(\i*\fcPreviewShiftX,\i*\fcPreviewShiftY)$); }%
  \end{pgfonlayer} 
}

\FCblock{fl}{$5\times$ FC blocks }{Out: $5\times$ regular, $|G|{=}4$}{F1}

\draw[->] (fl.south) -- (F1.north);
\node[ibox, right=4mm of F1, text width=14mm] (output) {output};
\draw[->] (F1) -- (output);

\end{tikzpicture}
\caption{Schematic layout of a traditional (non rotation equivariant) CNN with the same number of hidden layers as our \ECNN (shown in Fig.~\ref{fig:E2CNN}). We highlight that the standard CNN has the same number of Convolution blocks and Fully-Connected blocks as the \ECNN; the only difference is the presence here of a standard 2D Convolution Layer as opposed to the Equivariant Convolution Layer (R2Conv) for the \ECNN.} \label{CNN}
\end{figure}
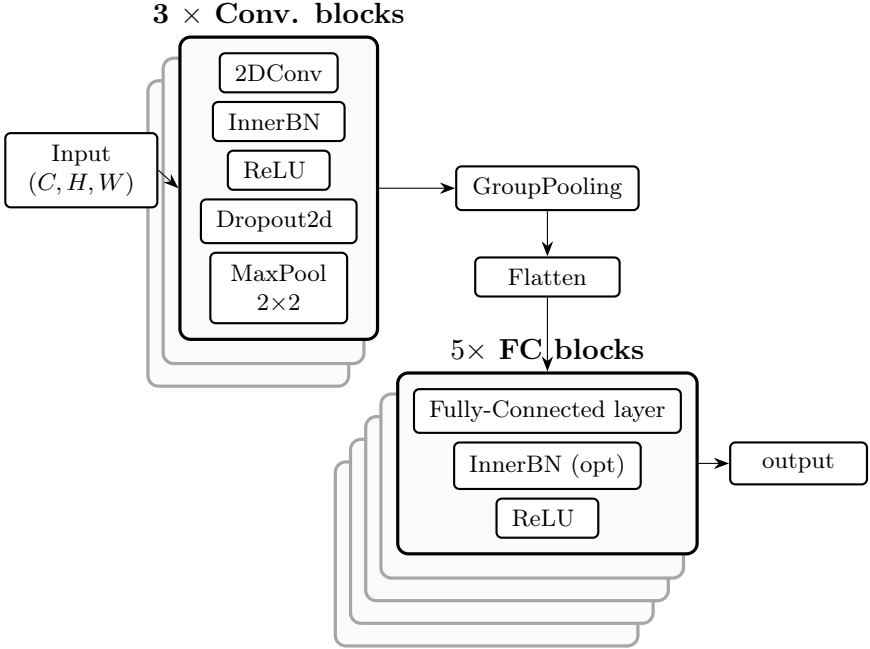

In this section, we compare the generalization capabilities of the \ECNN to those of a standard CNN (with the architecture shown in Fig. \ref{CNN}) --when both are trained on the original training dataset described in section \ref{sec:dataset}-- against rotations of the simulated galaxies in the test set.  This is a crucial test to prove that the potential random alignment of galaxies within our large yet limited training sample does not impact significantly our reconstruction results.

We build two versions of our test sample by rotating each galaxy's image channel by $\pi$ and $\pi/2$, respectively. 
To each rotated test sample we apply, separately, \ECNN and the standard CNN and infer the DM profile.
We then compare --for each rotated test set, and for each of the two architectures, separately-- the results against the original non--rotated test sample.

\begin{figure}[h]
    \centering
    \includegraphics[width=0.45\linewidth]{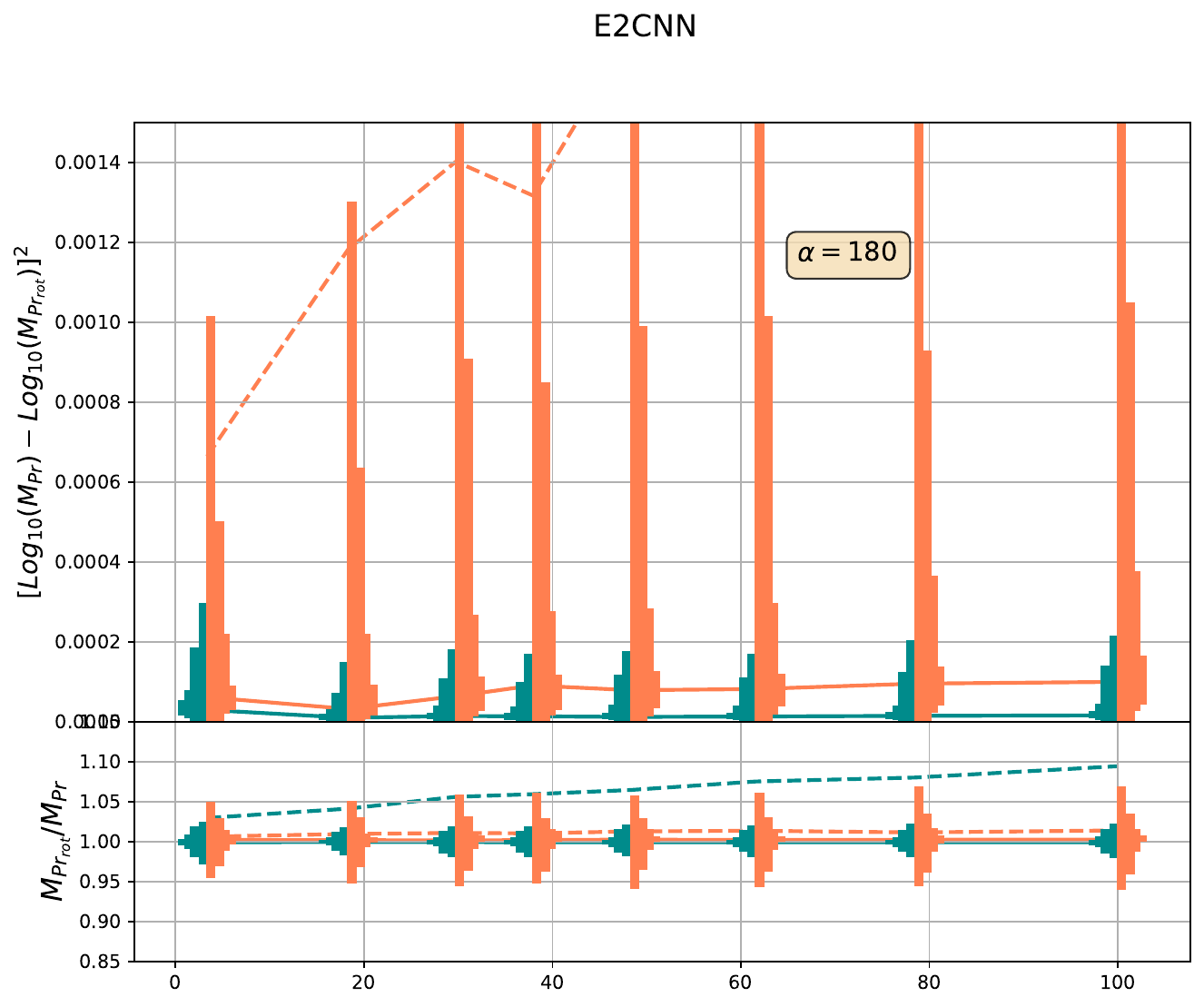}
    \includegraphics[width=0.45\linewidth]{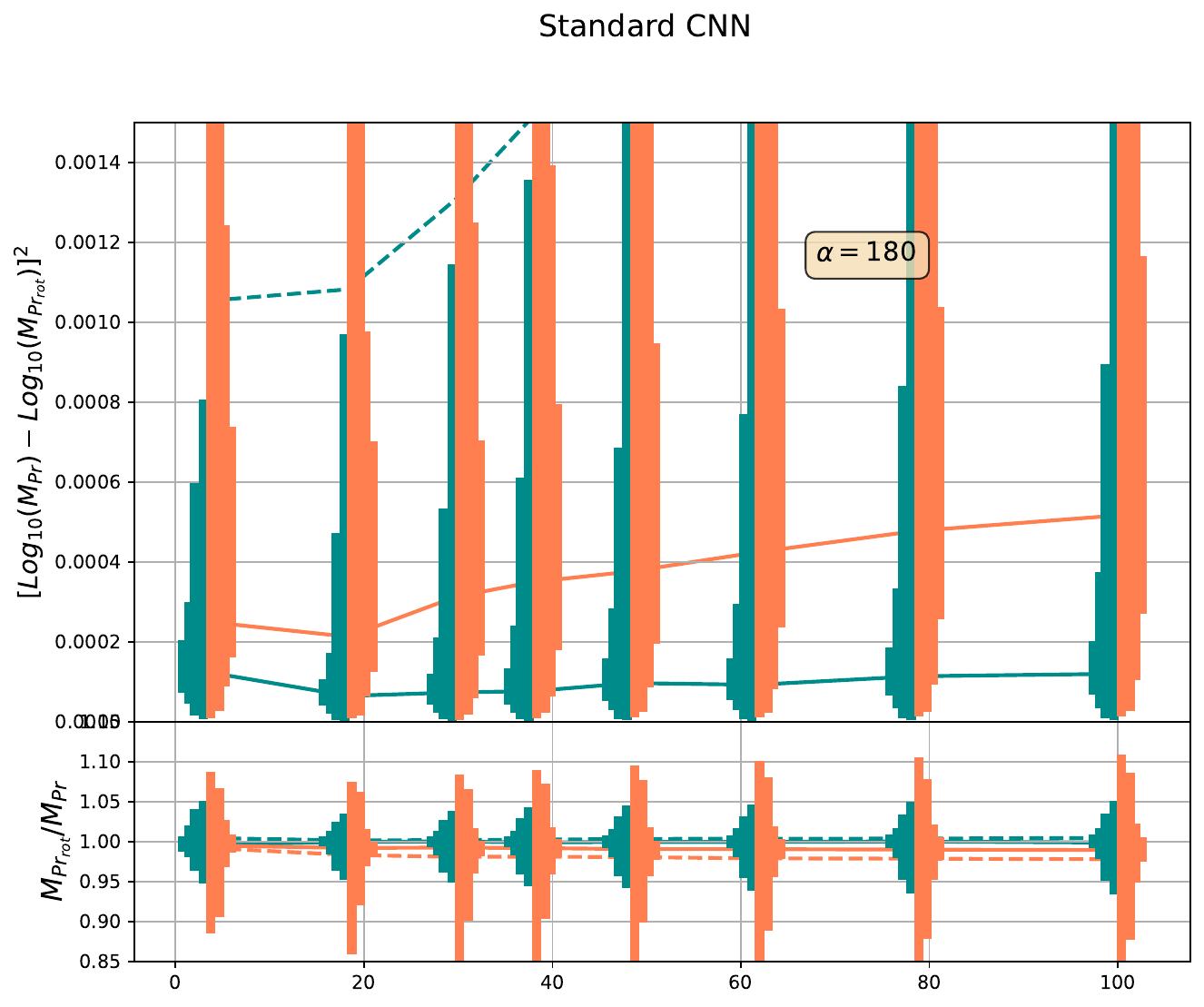}
    \includegraphics[width=0.45\linewidth]{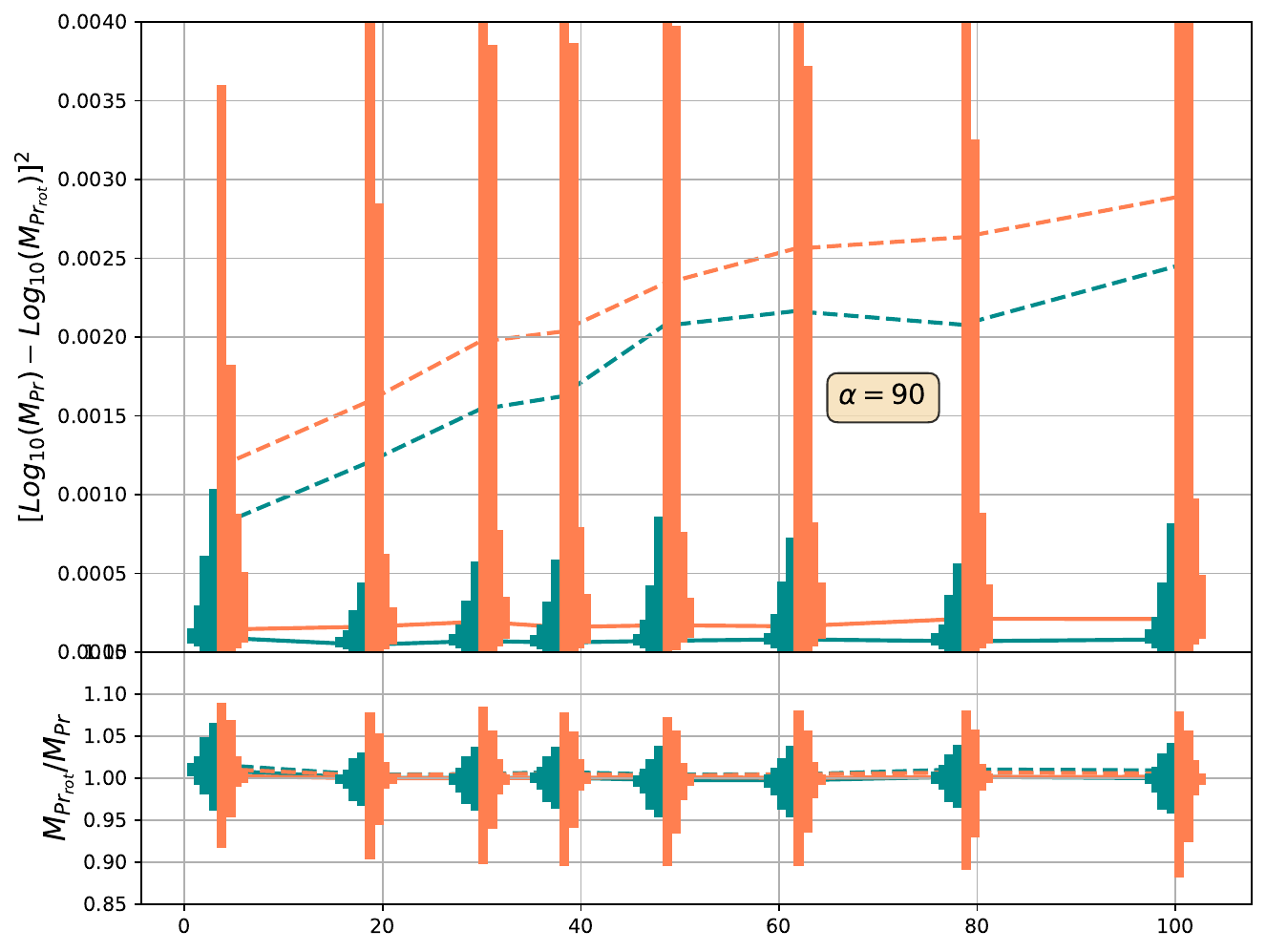}
    \includegraphics[width=0.45\linewidth]{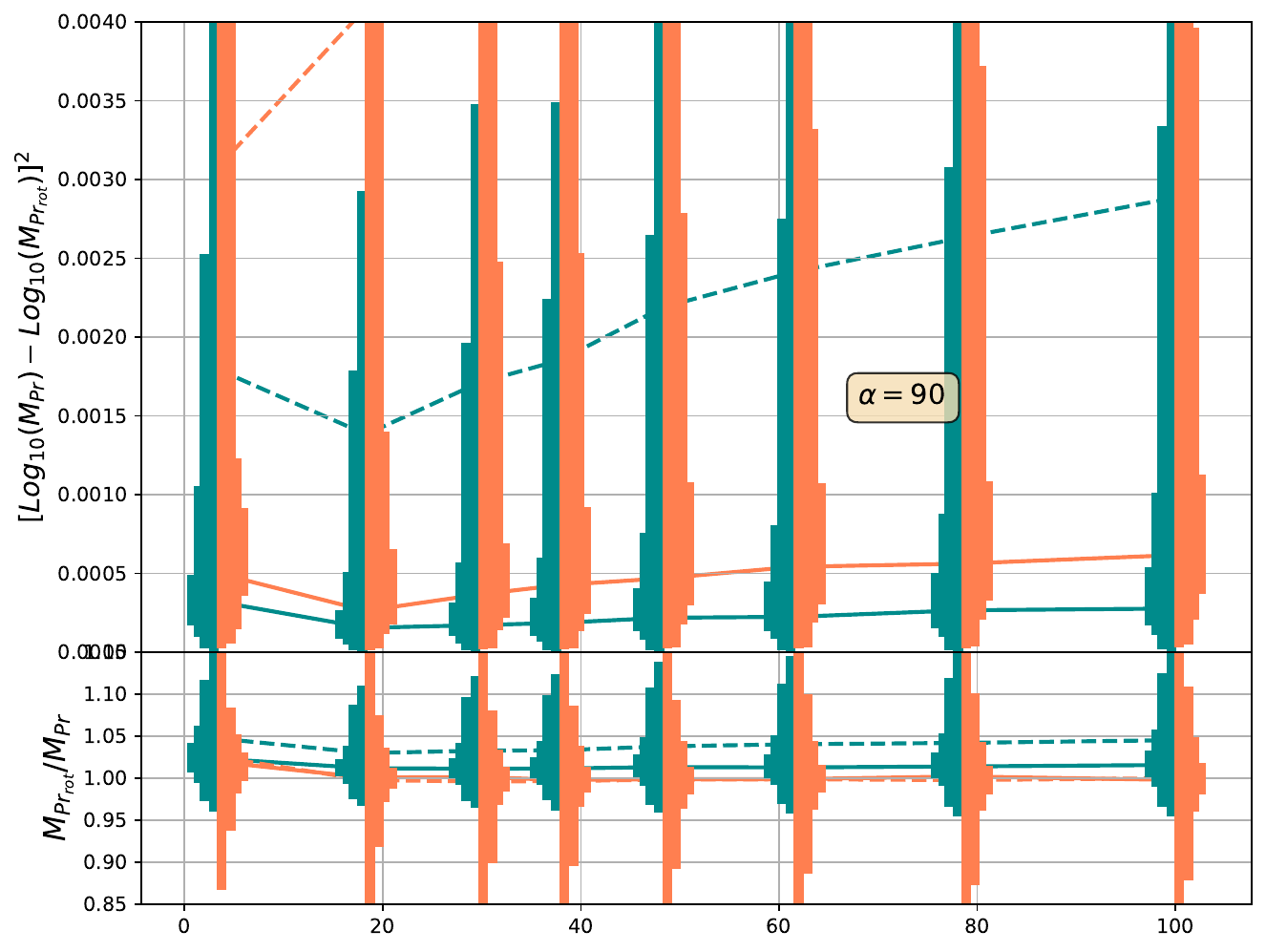}
    \includegraphics[width=0.13\linewidth]{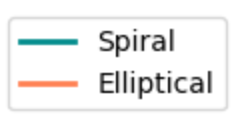}
    
    \caption{Error distribution of the inferred mass profile of the rotated vs non--rotated test sample. Each panel shows the comparison between the distribution with the specified architecture (for a given test set rotation angle), vs the non--rotated case. Left panels show \ECNN architecture results, right panels standard CNNs. Different rows pertain to different rotation angles: $\pi$, $\pi/2$, respectively. In the figure we show the results for spiral galaxies in green and for elliptical ones in orange, as reported in the legend.}

    \label{fig:rotation180}
\end{figure}

The results are shown in Figs. \ref{fig:rotation180}, where each panel shows violin plots comparing the specific test case  to the original, non-rotated case, {\it thus presenting a MSE relative to the rotation operation}. Panels on the left are for the \ECNN while those on the right refer to the standard CNN. 
It stands out clearly that \ECNN always performs significantly better than the traditional CNN in the case of rotated galaxies, in particular within numerical noise.

This is consequence of the fact that the implemented \ECNN architecture preserves equivariance for rotation angles that are a multiple of $\pi/2$, while the CNN does not. 

\newpage

\newpage


\bibliographystyle{JHEP}
\bibliography{biblio.bib}

\end{document}